\documentclass[11pt]{article} 

\usepackage[utf8]{inputenc} 

\usepackage{graphicx} 
\usepackage{amsthm}

\usepackage{geometry} 
\usepackage{graphicx} 

\usepackage{booktabs} 
\usepackage{array} 
\usepackage{paralist} 
\usepackage{verbatim} 
\usepackage{subfig} 
\usepackage{amsmath, amsthm, amssymb}
\usepackage{authblk}

\newtheorem{theorem}{Theorem}[section]
\newtheorem{lemma}{Lemma}[section]
\newtheorem{corollary}{Corollary}[section]
\newtheorem{definition}{Definition}[section]
\newcommand*{\QED}{\hfill\ensuremath{\square}}

\usepackage{fancyhdr} 
\usepackage{sectsty}
\allsectionsfont{\sffamily\mdseries\upshape} 

\usepackage[nottoc,notlof,notlot]{tocbibind} 
\usepackage[titles,subfigure]{tocloft} 

\title{Modelling Breakage-Fusion-Bridge Cycles as a Stochastic Paper Folding Process}
\author[1,2]{Greenman CD*}
\author[3]{Cooke SL}
\author[3]{Marshall J}
\author[3]{Stratton MR}
\author[3,4]{Campbell PJ}
\affil[1]{Department of Computing, University of East Anglia, Norwich, UK}
\affil[2]{The Genome Analysis Centre, Norwich Research Park, Norwich, UK}
\affil[3]{Cancer Genome Project, Wellcome Trust Sanger Institute, Hinxton, UK}
\affil[4]{Department of Haematology, University of Cambridge, Cambridge, UK}

\date{} 

\begin{document}
\maketitle

\allowdisplaybreaks

\begin{abstract}
Breakage-Fusion-Bridge cycles in cancer arise when a broken segment of DNA is duplicated and an end from each copy joined together. This structure then `unfolds' into a new piece of palindromic DNA. This is one mechanism responsible for the localised amplicons observed in cancer genome data. The process has parallels with paper folding sequences that arise when a piece of paper is folded several times and then unfolded. Here we adapt such methods to study the breakage-fusion-bridge structures in detail. We firstly consider discrete representations of this space with 2-d trees to demonstrate that there are $2^{\frac{n(n-1)}{2}}$ qualitatively distinct evolutions involving $n$ breakage-fusion-bridge cycles. Secondly we consider the stochastic nature of the fold positions, to determine evolution likelihoods, and also describe how amplicons become localised. Finally we highlight these methods by inferring the evolution of breakage-fusion-bridge cycles with data from primary tissue cancer samples.
\end{abstract}

\let\thefootnote\relax\footnote{*Corresponding Author}


\section{Introduction}

\begin{figure}[t!]
\centering
\includegraphics[height=140mm,width=150mm]{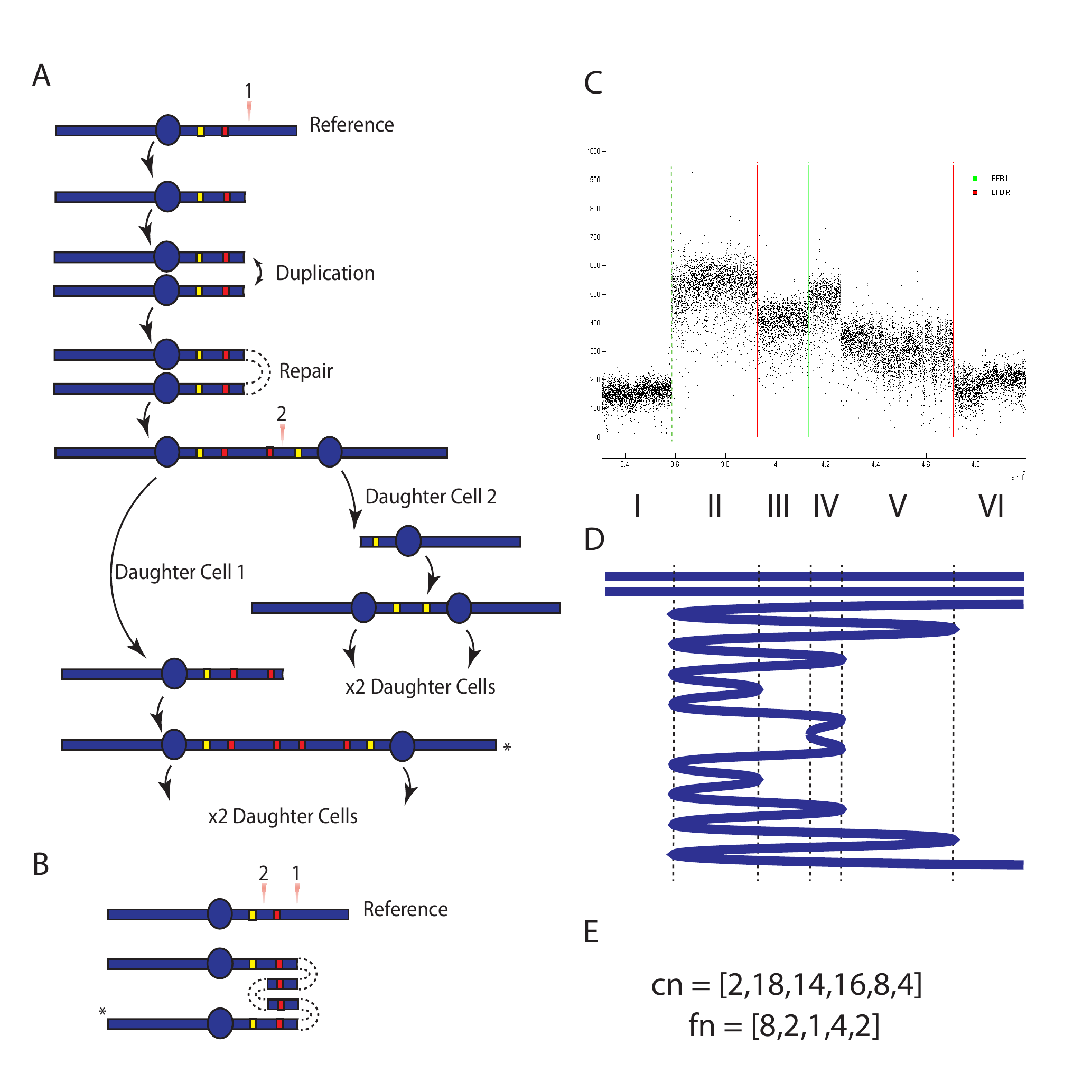}
\caption{The BFB Process. A) A representation of a chromosome, the circle being a centromere, the red and yellow markers hypothetical genes duplicated and deleted throughout this process. A DNA break (at position of orange triangles) followed by duplication and repair results in a palindromic chromosome with two centromeres. Spindles grab each centromere and contract during cell division resulting in another break and the cycle continues. B) The BFB product (*) is folded relative to the original reference genome. C) A typical amplicon formed through a BFB process. The horizontal axis is genomic position, the vertical axis is read depth. Vertical lines indicated detected BFBs. D) The predicted BFB folding pattern. E) The copy number profile; $cn$ counts the number of genomic copies in each region and $fn$ counts the number of folds at each fold loci.}
\label{Amplicon}
\end{figure}

Breakage-Fusion-Bridge (BFB) cycles potentially arise whenever a stretch of DNA is broken and a cell division cycle takes place. The first stage in this division process is DNA replication, where duplication will take place up to the DNA break, leaving two exposed ends. Prior to cell division, the cells checkpoint machinery will look for mistakes and the two exposed ends may be erroneously joined together by double stranded break repair. This results in a palindromic sequence of DNA, often containing a duplicated centromere (see Figure \ref{Amplicon}A). Spindles then attach to centromeres, which then contract during cell division to pull a chromosome into each daughter cell. However, if each centromere of this dicentric chromosome is to successfully relocate to distinct daughter cells, the DNA between the centromeres has to snap, and so each daughter cell will have a centromere on a DNA segment with an exposed (broken) end. This process of duplication, end repair and DNA breaking during can then continue with each cell division and the process repeat itself in a cascade of BFB cycles.

The process is unlikely to continue indefinitely because eventually repair machinery will attach exposed ends to other portions of the genome to produce a translocation, for example, or telomerase may cap the end to produce a somatic telomere. However, this process of repeatedly duplicating, repairing and unfolding is known to produce complex rearrangements of the original genomic assembly, and are frequently observed in cancer genomes \cite{BFB1, Bignell1}. The complexities of these rearrangements have also been examined algebraically in \cite{Kinsella}. 

These genomic rearrangements closely resemble paper folding operations in origami where paper is repeatedly folded in upward and downward directions. When the paper is unraveled, we obtain a series of troughs and peaks which can be represented as a binary sequence. These sequences can be recursively generated and serve as examples of automaton \cite{Automata}, which gives us a starting point to represent BFB processes.

There are some key differences to note however. Whilst paper is intrinsically the same material at all positions, DNA is composed of a variable sequence of nucleotides and subsequently is identifiably unique along its length \emph{prior} to the BFB duplication process (DNA repeats are ignored). We can thus label each point along the original DNA sequence with its genomic position and consider how these labels are duplicated in the BFB process. By comparing the positions on the BFB product with the original \emph{reference} sequence, we can fold the BFB product so the same labels (i.e. reference positions) are vertically aligned, such as in Figure \ref{Amplicon}B, where three folds are required. Note that these folds are located at precisely the two reference positions of DNA repair in the BFB cycles. The term \emph{fold} and the folded structure relative to the reference will be used in the majority of the work. We  will also refer to the stretch of DNA between two consecutive folds as a \emph{segment}. 

This representation mirrors that observed experimentally. In Figure \ref{Amplicon}C, for example, we have the results of some next generation sequencing. The horizontal axis represents the reference position, and the vertical axis the experimental signal (sequence read depth \cite{Pleasance1, Pleasance2}). We see this signal is relatively constant within each of six regions $I,II,...,VI$. The changes in signal where regions meet also coincides with positions of DNA folds detected with next generation sequencing; the green and red vertical lines indicating folds pointing left and right, respectively. Collectively, these results are indicative of a sequence of BFB cycles, and we will later infer the likely underlying folding structure, the prediction indicated in Figure \ref{Amplicon}D.

This inference relies in part on the linear relationship between the experimental signal and the number of copies of DNA `folded' across a given reference region, the \emph{copy number profile}, summarized in Figure \ref{Amplicon}E. The top \emph{copy number} vector $cn$ counts the number of copies of each region in the final structure, and \emph{fold number} vector $fn$ counts the number of copies of each fold. For this prediction, we see that region $II$ (with the highest signal) has a predicted copy number of $18$; $16$ copies in the BFB structure and $2$ from other copies of the original chromosome. The fold numbers at the left and right side of this region are $2$ and $8$, respectively, reflecting the number of times DNA reverses direction relative to the reference. Note that as we move from region $I$ to $II$, the copy number changes by $16$, and the fold number is half of this, $8$. This is because each fold accounts for two genomic copies.

These data constitute an example of an `amplicon', which are frequently observed in cancer genome data. These are clusters of rearrangements with a high signal in the reference genome, indicating an abnormally large number of copies are present in the cancer genome. These `amplified' regions are usually restricted to a few megabases of DNA, a small proportion of a typical human chromosome. The BFB process is one mechanism by which these events can arise \cite{BFB1, Bignell1}. Next generation sequencing technologies mean we can now visualize these events in great detail, producing extensive catalogs of the mutations involved \cite{Pleasance1}, \cite{Pleasance2}, from which the etiology of these events can then be investigated \cite{Greenman1}, \cite{Raphael2}.

In this work we consider several interesting questions that naturally arise from the processes we are considering and the data they produce. Firstly, consider the problem of how best to represent this process. It is both  discrete, in terms of the type of folded structures that can arise, and continuous, in terms of the reference positions of the folds. By introducing a discrete representation of the BFB process, we provide a coherent representation of the genomic conformations that can arise in BFB `space'. Furthermore, this structure allows us to measure the size of this space, proving that there are $2^{\frac{n(n-1)}{2}}$ qualitatively distinct evolutions given $n$ BFB cycles. We also model the process stochastically. This allow us to reconstruct the most probable evolution of BFB processes underlying any observed amplicon. Furthermore, this provides some understanding into why amplicons are so localised in the genome.  

The paper is arranged as follows. The next section considers how to discretely represent the process as an iteration on algebraic words, and how this iteration can be inverted, extending some ideas in \cite{Kinsella}. This enables the BFB process to be identified from the final DNA sequence (but not the copy number profile, which contains less information). We then introduce the more compact \emph{BFB sequence} which is a discrete way of labeling the BFB process. This is non-unique, and each BFB sequence produces a partially ordered set (poset) of possible BFB processes, each with a range of possible copy number profiles. The subsequent section introduces methods to produce and count the BFBs contained within each poset and hence determine how the size of the space of BFB products grows with the number of BFB cycles that have taken place. We then consider the stochastic nature of the fold positions, and the impact this has on the likelihood of observing possible BFB structures. Applications of these methods and the difficulties of dealing with real data are then explored in more detail. Concluding remarks complete the paper. All proofs can be found in the Appendix.


\section{Word Representations of BFB Processes}

We now consider algebraic representations of the BFB process, both in a forward direction, reflecting the evolution of the BFB process, and backward, indicating how to unravel a BFB folded structure, reverse the process, and infer the events that have taken place.


\subsection{The Forward Process}

The BFB process can be described as an iteration scheme on a word of symbols. This follows ideas from paper folding sequences, where the binary letters of words represent peaks and troughs that run along an unfolded piece of paper formed from a series of folding operations. These words can be constructed by an iteration of word operators each of which depend upon whether the folding action was up or down \cite{Automata}.

For BFBs we have to generalize this somewhat. Firstly, binary sequences prove inadequate, and there are two symbolic `languages' that we draw from; one is in terms of the (reference) positions of folds, the other is in terms of the regions of the reference genome that bridge these positions. The latter representation was considered in \cite{Kinsella}.

\begin{figure}[t!]
\centering
\includegraphics[height=140mm,width=140mm]{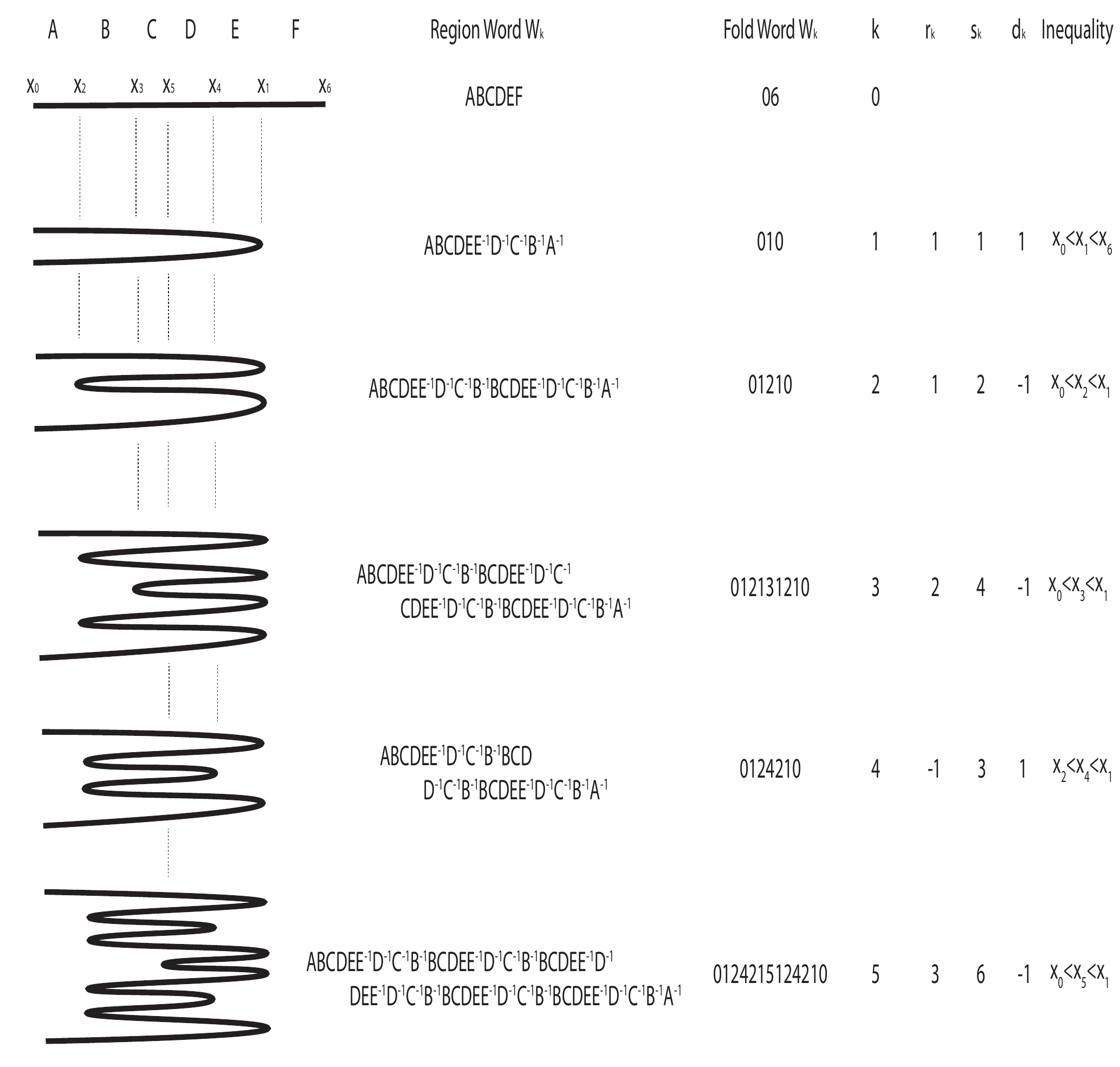}
\caption{The first column indicates a sequence of five BFBs. The second and third columns gives region and fold word representations. The fourth column indexes the BFB. The fifth column indicates the BFB sequence. The sixth and seventh columns are the cumulative representative sequence and the directional signature. The final column provides inequalities satisfied by the fold positions.}
\label{Intro}
\end{figure}

Consider the example in Figure \ref{Intro}. Here a segment of DNA undergoes a series of five BFB cycles (first column). This results in five folds positions which partition the reference genome into six regions, which we label $A$, $B$, $C$, $D$, $E$ and $F$, our first language. These are initially contiguous and we represent this as the \emph{region} word $ABCDEF$, as given in the second column. Our second language utilizes the reference positions of BFB folds, labeled $0,1,...,6$, which includes the ends of the original structure. Because no BFBs have yet taken place, we have the trivial \emph{0-fold} word $06$, representing the ends of the region, as indicated in the third column. The `\emph{0-}' indicates no folds have yet occurred.

We now implement the BFB process. Firstly, we have a break in our segment at the $1^{\emph{st}}$ fold position, $x_1$, separating regions $E$ and $F$. We suppose the DNA to the left of the break is duplicated and the right side is lost to a different daughter cell. The two duplicated ends at position $x_1$ are then stitched together to form one new structure. If we read the regions traversed through this new structure, we have word $ABCDEE^{-1}D^{-1}C^{-1}B^{-1}A^{-1}$, where a negative power indicates the region is in a reversed direction. Note that the sub-word $..EE^{-1}..$ essentially demarcates the right pointing fold at this position. Note also that the word is pseudo-palindromic; the letters are palindromic and the signs on the right hand side are opposite to those on the left. If we reverse the order of the symbols and change their signs we find the word is equal to its inverse, $W^{-1}=W$, which reflects the chromosome symmetry; if we turn the chromosome upside down, we end up with an identical structure. In the second language of folds we traverse the structure from one end reporting fold numbers when we reach a fold; this gives the simpler \emph{1-fold} word $010$. Unlike the region word representation, we do not have signs to indicate direction and note that as we traverse the structure, the folds alternate the direction they point. We will see below that both representations are equivalent.

We continue the process, the next fold occurring at position $x_2$, resulting in a structure with three folds and four segments, with corresponding word $ABCDEE^{-1}D^{-1}C^{-1}B^{-1}$ $BCDEE^{-1}D^{-1}C^{-1}B^{-1}A^{-1}$. This time the fold points to the left and the subword $..B^{-1}B..$ represents the fold between segments $A$ and $B$. In terms of folds we now have \emph{2-fold} word $01210$.

Note that we have two choices to construct the fold at reference position $x_2$. Considering the second structure formed, we can either break at the upper copy of $x_2$ and duplicate the DNA below, or break at the lower copy and duplicate the DNA above, the same product results. This symmetry is true in general, where we have the following. 

\begin{lemma}
\label{Symmetry}
If a BFB fold is positioned a length $l$ from one end of a BFB product, with the upper portion duplicated, the BFB product cannot be distinguished from that arising when a BFB fold positioned a length $l$ from the other end with the lower portion duplicated.
\QED
\end{lemma}

The palindromic nature of BFB products then means we always have two choices to place the fold position for any given product. In what follows, we always assume that we are duplicating DNA above the position of the fold, with respect to the representations drawn in Figure \ref{Intro}. Note that this only applies to a palindromic product and so does not apply to the very first BFB event, for which every fold position and duplication gives a unique product.

We now continue the process, iteratively building up the word. The next new fold is at position $x_3$, occurring after the third fold of the third product. We thus keep the word containing the first three folds $0121$, insert the new fold $3$, and add the first three folds in reverse order $1210$, resulting in \emph{3-fold} word $012131210$. This fold is then deleted by the fourth BFB; the corresponding fold is positioned at reference position $x_4$, occurring immediately after the second fold in the structure and we obtain \emph{4-fold} word  $0124210$, losing symbol $3$. Although five folds take place, the final conformation $0124215124210$ then only contains four fold numbers $1,2,4$ and $5$. This can happen generally, if the fold occurs in the upper half we lose the middle position, which must be the previous BFB location, and information is lost. Furthermore, if we simply implement the BFBs $1,2,4$ and $5$ we get the same product. We refer to this as a \emph{reduced} BFB set.

Note also that in our example the first BFB involved the loss of the segment to the right of $x_1$, resulting in two ends extending in a leftward direction relative to the reference. We have lost all copies of the rightwards ends after this step, and every subsequent BFB results in two ends that always point in a leftward direction. In this sense the direction of the first BFB event is special. We refer to this direction as the BFB \emph{parity} $p$, where $p=1$ (resp. $-1$) indicates the ends extend to the right (resp. left), relative to the reference.

We summarize the word representations of a BFB process as follows.

\begin{lemma}
The region word $W_i$ representation of the $i^\emph{th}$ step of $n$ BFB processes is constructed by taking the initial word $W_1=S_1S_2...S_nS_n^{-1}...S_1^{-1}$ for a parity $-1$ BFB, or $W_1=S_{n+1}^{-1}S_n^{-1}...S_2^{-1}S_2...S_{n+1}$ for parity $1$ and constructing the word $W_i$ recursively from $W_i=W_{i-1}(a_i)W_{i-1}(a_i)^{-1}$, where $a_i$ is the number of regions from the upper end of the ${i-1}^{th}$ BFB product that are replicated, and $W(j)$ truncates word $W$ to the first $j$ symbols.

For the fold word we have $W_1=010$ and recursion $W_i=W_{i-1}(b_i)iW_{i-1}(b_i)^{-1}$ where $b_i-1$ is the number of duplicated folds.
\QED
\end{lemma}

This gives us two representations for BFB structures. The region word is somewhat more cumbersome, but gives a readout of consecutive reference regions in the structure and as such provides a \emph{contig} of the underlying chromosome. Note also that counts of individual letters in the word also provide the vector $cn$ of the copy number profile. The fold word is the more efficient representation. Counts of distinct symbols in this word provide the counts $fn$ of the copy number profile. For example, the words associated with the final structure in Figure \ref{Intro} have ten occurrences of $B$, the number of copies of the second region, and four occurrences of $1$, the number of folds at position $x_1$. Not all integer sequences can arise as a BFB copy number profile; the range of possible copy number profiles $cn$ has been explored in detail in \cite{Kinsella}.


\subsection{The Reverse Process}

We now consider the opposite problem; given a BFB word, we need to reverse the process and identify the events that have taken place. This represents the typical inference problem in genomics, where we have the final structure of a genome and wish to reverse engineer the process to identify the evolutionary history. This can be achieved by identifying the unique BFB fold that demarcates the centre of the palindromic structure and undoing the duplication. For example, the BFB sequence of Figure \ref{Intro} resulted in fold word $W=0124215124210$. The centre fold $5$ is undone, leaving $012421$. This must arise from palindrome $0124210$ so we undo $4$ to leave $012$, which must derive from palindrome $01210$. Undoing $2$ and then $1$ completes the sequence and the evolution of folds is $1,2,4$ and then $5$. Note that we have reconstructed the reduced BFB set, not the full list; fold $3$, which was deleted by fold $4$, is not included. Note also that $1,2,4,5$ is precisely the order that the symbols first appear in the final word $W=0124215124210$. 

In general we have the following result (see Appendix)

\begin{theorem}
A fold word is a viable representation of a BFB process if and only if it can be reversed with the following algorithm. This identifies the unique \emph{reduced} BFB sequence responsible for the word.

\emph{STEP 1}: Take palindromic fold word $W=XnX^{-1}$ and undo fold $n$ to construct $X$. Output $n$.

\emph{STEP 2}: Identify the rightmost uniquely occurring symbol $m$ such that the fold word is $ZYmY^{-1}$ for some (possibly empty) subsequences $Z$ and $Y$ which do not contain $m$. Undo $m$ and contract to word $W=ZY$. Output $m$.

\emph{STEP 3}: If $W$ is empty the reduced evolution is the reverse of the output, else repeat \emph{STEP 2}.

For a viable BFB fold word $W$, the evolutionary order of BFBs is precisely the order that their corresponding fold number first appears in the word.
\label{Algorithm}
\QED
\end{theorem}

Thus we find that if we know the genomic sequence arising from BFBs, we can reverse the process and indicate the unique minimal sequence of BFBs that lead to that sequence. Unfortunately, experimental data does not always contain such detailed information. The copy number profile, for example, is a more typical experimental observable, indicating the number of times different regions are present, but not the order that they are present in the chromosomal structure. Furthermore, we have not considered the random nature of the process and in particular the different orders the fold positions can take. To help deal with these issues we next introduce a representation which captures the events that take place, rather than the sequence generated.


\section{BFBs as a Discrete Process}

Here we introduce a discrete representation of the BFB process. For every $n$-fold word we obtain a sequence of $n$ integers which proves more analytically tractible. This allows us to count the number of qualitatively distinct BFB evolutions for a set number of BFB cycles.

\subsection{BFB sequences}

Consider the evolution portrayed in Figure \ref{Intro}. After each BFB we have a folded structure with a finite set of DNA segments going forward and backward between fold positions relative to the reference. We shall construct a \emph{BFB sequence} ${r_n}$ to represent these structures as follows. In order to specify a BFB event we have to indicate where the next fold is positioned on the current structure. The symmetry of the process (Lemma \ref{Symmetry}) means we can specify that the duplication will always occur from one end, so we choose the top end of each structure as presented in Figure \ref{Intro}. We then have to indicate which segment the next BFB fold will occur upon. For reasons described below, the segment immediately after the mid point is labeled $1$ and the labels of segments before or after are obtained by counting backward or forwards along the structure, respectively. The value $r_n$ is then the label for the segment containing the $n^{\emph{th}}$ BFB fold.

Thus for the example of Figure \ref{Intro} we start with one segment. The first BFB occurs on this segment so we trivially write $r_1=1$. This produces two segments, and so two choices for the location of the next BFB fold. In our example this occurs on the edge below the midpoint, so we have $r_2=1$, producing four segments. The next BFB occurs on the last segment, two segments after the midpoint, so $r_3=2$. The next BFB loses the $3^{\emph{rd}}$ BFB fold, occuring two segments before the midpoint, so counting back from $1$, we have $r_4=-1$. The final BFB gives us $r_5=3$ so we have \emph{BFB sequence} $r=[1,1,2,-1,3]$, as indicated in the fifth column of Figure \ref{Intro}.

We noted previously that because the $3^{\emph{rd}}$ BFB is deleted by the $4^\emph{th}$, the end product can be obtained by simply implementing the undeleted BFBs. Note that the $4^\emph{th}$ BFB fold can be positioned on the third segment after the midpoint of the structure arising from the $2^\emph{nd}$ BFB event. We can thus equally represent the final structure with \emph{reduced BFB sequence} $r=[1,1,{\bf 1},3]$. Note that this can be derived from the full sequence $r=[1,1,2,-1,3]$ by absorbing the negative value $-1$ into the preceding value $2$, giving new (emboldened) value $1$.

Consider the cumulative sum of the full sequence, ${\bf s}=[1,2,4,3,6]$. We have two observations. Firstly, note that $s_n$ indicates the number of segments into the structure that we first encounter a copy of folds arising from the $n^{\emph{th}}$ BFB. For example, the fold from the $5^{\emph{th}}$ BFB, is $6$ segments into the final structure. Secondly, values $p(-1)^{s_n}$, where $p=-1$ is the parity, provide a \emph{direction signature}, ${\bf d}=[1,-1,-1,1,-1]$. Each number $d_n$ gives the direction of all copies of the $n^{\emph{th}}$ BFB fold, relative to the reference. For example, all copies of the fold from the $2^{\emph{nd}}$ BFB, at position $x_2$, point to the left ($d_2=-1$). Note that all these signs would be flipped if the original BFB had reversed parity, facing the opposite way.

Representing the process this way thus enables us to capture some nice properties. We summarize this as follows.

\begin{theorem}
Any $m$-fold word representing a BFB process has an equivalent representation with a sequence of $m$ integers $\{r_n\}$ that satisfies $-s_n < r_{n+1} \leq s_n$ where $s_n=\sum_{i=1}^{n}r_i>0$. Values $s_n$ indicate the number of segments into the structure that we first encounter a copy of folds arising from the $n^{\emph{th}}$ BFB cycle. Each term $r_n$ in the sequence represents a BFB event. Negative terms indicate deletion of previous BFBs which shorten the structure. The \emph{reduced representation} contains strictly positive values and is obtained by combining $r_{n-1}$ and $r_n$ into the single term $r_{n-1}+r_n$ whenever $r_n \leq 0$. This is repeated until only positive terms remain. The number of reductions using $r_n$ equals the number of BFB events deleted by the $n^{th}$ event. The direction of all copies of folds from the $n^{th}$ BFB is given by $d_n=p(-1)^{s_{n}}$, where $p$ is the parity of the initial BFB.
\label{BFB_Seq}
\QED
\end{theorem}

\begin{figure}[t!]
\centering
\includegraphics[height=90mm,width=140mm]{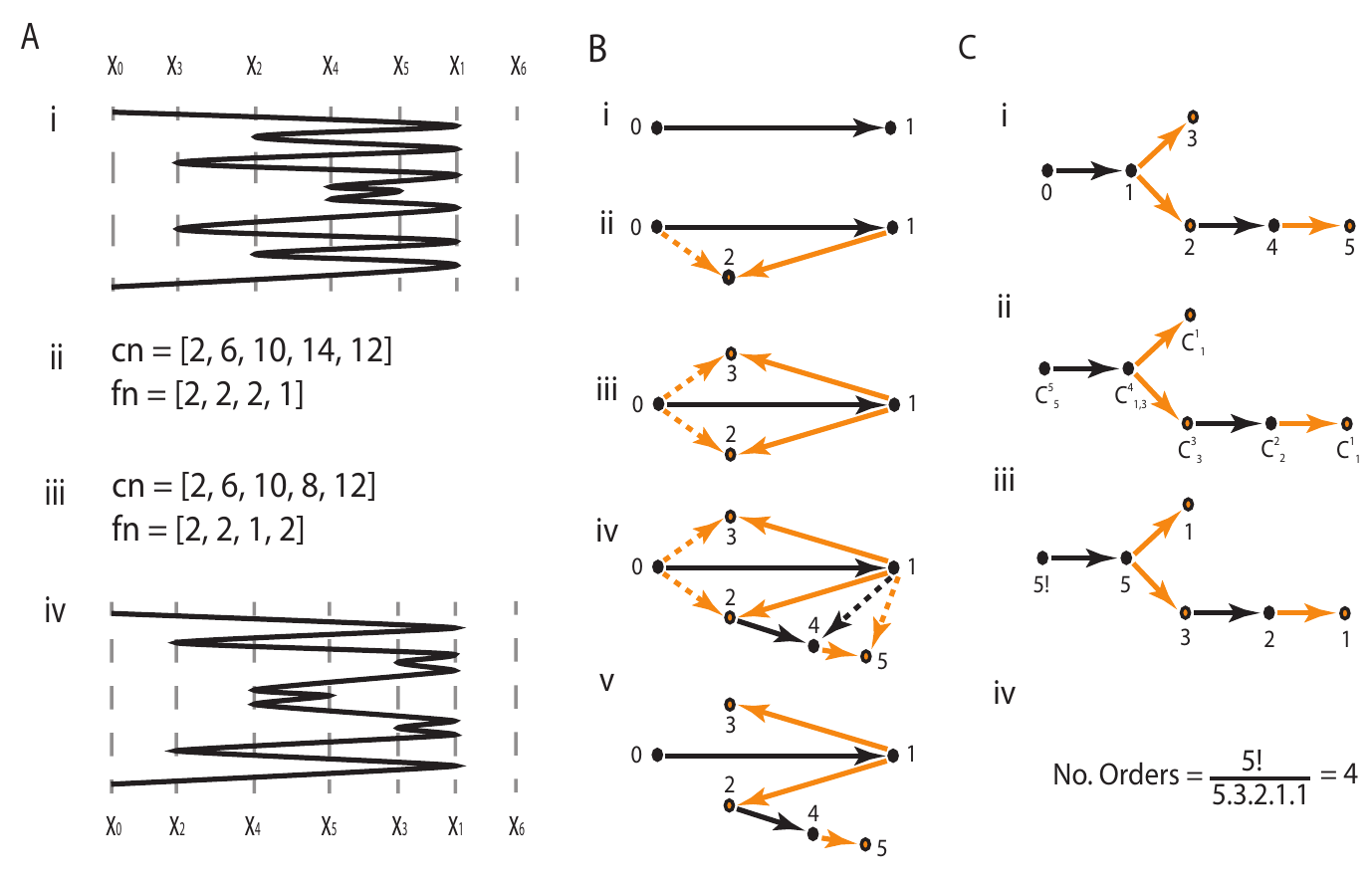}
\caption{Poset of folding structures for a given BFB sequence. In \emph{Ai,iv} we see two possible arrangements resulting from reduced representation $[1,1,2,2,1]$. \emph{ii,iii} gives the copy number profiles of each. In \emph{B} we see the poset graph construction representing the possible orders of positions. Nodes represent folds, edges represent inequalities between fold positions in the reference. Solid and dashed edges indicate \emph{major} and \emph{minor} edges, respectively. Black and orange edges indicate \emph{plain} and \emph{flipped} edges. Trees in \emph{C} indicate how Lemma \ref{PosetReduction} is used to count the number of possible orders in the Poset.}
\label{Posets}
\end{figure}

We now have a way of representing BFBs. We will see that each representation can account for many different BFB structures. The two structures given in Figure \ref{Posets}Ai,iv are both represented by sequence $[1,1,2,2,1]$, for example. We build this example sequentially. We start with a single segment, trivially represented by \emph{order} word $W_0=06$, which undergoes a BFB at position $x_1$, where $x_0<x_1<x_6$. This loses end $x_6$, which, reporting folds as we read through the structure, we associate with word $W_1=010$. The next BFB fold occurs on the segment after the midpoint, which we represent as word $W_2=01210$. The fold occurs at some position $x_2$ where $x_0<x_2<x_1$. The next segment occurs two segments after the midpoint, where $r_3=2$ and $s_3=4$. This is represented by word $W_3=W_2(s_3)3W_2(s_3)^{-1}=012131210$  ($W(m)$ represents the first $m$ letters of word $W$, and the negative power reverse the order) and we find that $x_0<x_3<x_1$, or $x_{W_{2,s_3+1}}<x_3<x_{W_{2,s_3}}$, where $W_{k,n}$ is the $n^{\emph{th}}$ letter of word $W_k$. We thus find that there are several order relationships on the reference positions of the BFB; we have a \emph{partially ordered set} (poset).

The general situation is described in the following result.

\begin{lemma}
\label{PosetDefine}
Define $W_n$ to be the word built with recurrence relation $W_n=W_{n-1}(s_n)nW_{n-1}(s_n)^{-1}$, initialised with $W_1=010$. If $x_n$ is the reference position of the $n^{th}$ BFB, and $d_n$ is the direction of Theorem \ref{BFB_Seq}, then we find that the following partial orderings apply to the positions of the BFB folds:

$\left\{
\begin{array}{l l l}
d_n = -1 & => & x_{W_{n-1,s_n+1}}<x_n< x_{W_{n-1,s_n}}\\
d_n = 1  & => & x_{W_{n-1,s_n}}<x_n< x_{W_{n-1,s_n+1}}\\
\end{array}
\right.$
\QED
\end{lemma}

For example $[1,1,2,2,1]$ we thus obtain restrictions $0<x_2<x_1$, $0<x_3<x_1$, $x_2<x_4<x_1$ and $x_4<x_5<x_1$. There are several different orders that satisfy these criterion, Figure \ref{Posets}Ai,iv being two such examples, where i has order $x_0<x_3<x_2<x_4<x_5<x_1$ and iv has order $x_0<x_2<x_4<x_5<x_3<x_1$. Note that this reordering of the fold positions induces a permutation in the fold numbers $fn=[2,2,2,1] \rightarrow [2,2,1,2]$, which in turn alters the copy number profiles (Figures \ref{Posets}Aii,iii). The copy number profiles for a given BFB sequence will thus be distinct in general unless the permutation permutes identical fold numbers.

It is natural to attempt to count and construct the different orders we get for a single BFB sequence. We do this with the aid of a 2-d tree construct; a special kind of directed graph that generalizes the notion of a tree, as exemplified in Figure \ref{Posets}B. When constructing a standard rooted tree, we can build from the root, recursively extending the tree with a single node and edge from a node that is already present. A 2-d tree differs in this respect in that once we have one edge and two nodes, each new node has two \emph{parent} nodes already present. Two edges are then constructed from these two nodes to the new node \cite{2dTree}.

We construct a 2d-tree as follows. Each new node represents a BFB cycle, with label $n$ representing the $n^{th}$ fold. Edges represent the ordering. When the $n^{th}$ fold is formed, it is positioned on a segment between two pre-existing folds from positions $x_a$ to $x_b$, where we have $a<b$, without loss of generality. We then construct two directed edges from nodes $a$ and $b$ to $n$. For example, the second fold in Figure \ref{Posets}Ai has position $x_2$ with $x_0<x_2<x_1$, we thus construct two edges from nodes numbered $0$ and $1$ to a node labeled $2$ as given in Figure \ref{Posets}Bii. 

We next introduce some classes for these edges.

\begin{definition}
Each pair of edges introduced during the 2d-tree construction consists of a \emph{major} and \emph{minor} edge. The \emph{major} edge (represented as solid edges in figures), extends from the node with greater value $b$, and the \emph{minor} edge (dashed), extends from the other node labeled $a$. 
\end{definition}

\begin{definition}
Each pair of edges and daughter node are either \emph{plain} or \emph{flipped}. If the word $W_{n-1}$ prior to the formation of fold $n$ changes from $..ab..$ to $..an..$  ($b>a$), the two edges and daughter node are \emph{plain} (black), if they change from $..ba..$ to $..bn..$, they are \emph{flipped} (orange). 
\end{definition}

For example, in Figure \ref{Posets}Biv we see node numbered $4$ extending from nodes $1$ and $2$. The major edge (solid) then extends from the larger source node numbered $2$. This node represents the introduction of the $4^{th}$ fold where word $01213{\bf 12}10$ becomes $012131{\bf 4}131210$. Because the two source nodes are increasing ${\bf 12}$ in the word, both edges and daughter node are termed plain (black). Conversely, the $5^{th}$ fold arises when $012131{\bf 41}31210$ becomes $0121314{\bf 5}4131210$ so the edges and node are flipped (orange).

The following observation is important when we later consider the number of possible evolutions from a fixed number of BFB cycles.

\begin{lemma}
Consider a node with value $n$ ($n>b>a$) constructed such that the major and minor edges are attached to nodes with values $b$ and $a$, respectively. Then:

{\bf I} If node $n$ is plain (black), any new node with major edge connected to $n$ has a minor edge connected to $n$'s minor parent $a$.

{\bf II} If node $n$ is flipped (orange), any new node with major edge connected to $n$ has a minor edge connected to $n$'s major parent $b$.
\QED
\label{MinorParent}
\end{lemma}

Thus consider Figure \ref{Posets}Biv, for example. Node $4$ extends from flipped node $2$ and so has major edge connected to $2$ and minor edge connected to $2$'s major node $1$. Node $5$ extends from plain node $4$ and so has major edge connected to $4$ and minor edge connected to $4$'s minor node, $1$.

This construction is termed the \emph{2-d Poset Tree}, $P$. Now, the parental nodes labeled $a$ and $b$ must correspond to a segment in the folded structure, so their positions are ordered relative to each other. The introduction of the $n^{th}$ node is ordered relative to both. Because the node labeled $b$ is already ordered relative to $a$, we only need consider the order of $n$ relative to $b$. That is, the major edges describe the ordering information for the fold positions. By ignoring the minor edges the 2d-tree construction then becomes a standard tree construction, such as \ref{Posets}Bv. This is termed the \emph{Order tree}, $T$.

We now find that any pair of nodes on the same tree branch correspond to folds with fixed relative positions, whereas nodes on distinct branches correspond to folds that are not ordered relative to each other. If we have two branches of $k_1$ and $k_2$ nodes descending from the same parental node, we find that there are $^{k_1+k_2}C_{k_1,k_2}$ ways of intercalating their positions. If there are already $o_1$ and $o_2$ different possible orders for each branch we then find $o_1o_2 ^{k_1+k_2}C_{k_1,k_2}$ possible orderings. We thus find that if we place $^kC_{k_1,k_2,...,k_B}=\frac{k!}{k_1!k_2!...k_B!}$ at each node, where $k$ is the total number of daughter nodes and $k_b$ denote the number of daughter nodes down branch $b \in \{1,2,...,B\}$, the number of orders is obtained by simply multiply these combinatorial terms together, such as in Figure \ref{Posets}Cii. Combinatorial terms at each end of an edge will then largely cancel (Figure \ref{Posets}Ciii) and we are left with the numerator $n!$ at the root node, where there are $n+1$ nodes in total, and denominators $n_b$ which count the number of daughter nodes, plus one, for each node $b$. We then see in Figure \ref{Posets}Ciii,iv that BFB sequence $[1,1,2,2,1]$, corresponding to word $012131454131210$, has $4$ possible structures. Summarizing, we have the following.

\begin{lemma}
Let $T$ denote the order tree of a 2-d poset tree $P$ deriving from a fold word or BFB sequence on $n$ BFB cycles. Tree $T$ then has $n+1$ nodes. If and each node below the root has a label $m_b$ counting the number of daughter nodes plus one, then the number of orders is given by $O(T)=\frac{n!}{\prod_b m_b}$.
\label{PosetReduction}
\QED
\end{lemma}

\subsection{The Size of BFB Space}

We now consider the question of how many different BFB sequences are possible, both for the case of reduced and non-reduced sequences. Although closed forms for these counts would seem intractable, we can derive counts recursively, where we have the following result.

\begin{lemma}
\label{BFB_Counts}
Let $v_1=w_1=(1,0,0,...)$ denote the infinite vector with single unit entry. We construct general vectors $v_n$ with components $v_{n,m}$ through the recursive relation $v_{n+1,m}=\sum_{k=\lfloor\frac{m+1}{2}\rfloor}^{m-1}v_{n,k}$. Then the number of reduced BFB sequences of length $n$ is $\sum_{m}v_{n,m}$. Applying the recursion $w_{n+1,m}=\sum_{k=\lfloor\frac{m+1}{2}\rfloor}^{\infty}w_{n,k}$ yields the number of full BFB sequences of length $n$ as $\sum_{m}w_{n,m}$.
\label{BFB_Count}
\QED
\end{lemma}

\begin{table}
\begin{center}
\scalebox{0.9}{%
  \begin{tabular}{| c || c | c | c | c | c | c | c| c| c| c |}
    \hline
    \text{BFBs}					& 1 & 2 & 3 & 4   & 5       & 6 & \dots & 10\\ \hline \hline
    \text{Reduced Sequences} 			& 1 & 1 & 2 & 7   & 41     & 397 &  & 627,340,987\\ \hline
    \text{Reduced Copy Number Profiles} 	& 1 & 1 & 3 & 19 & 247   & 6445 &  & - \\ \hline
    \text{Reduced Evolutions} 			& 1 & 1 & 3 & 21 & 315   & 9765 &  & 10,180,699,028,325 \\ \hline \hline
    \text{Full Sequences} 				& 1 & 2 & 6 & 26 & 166   & 1626 & & 2,290,267,226\\ \hline
    \text{Full Copy Number Profiles} 		& 1 & 2 & 5 & 24 & 271   & 6716 & & - \\ \hline
    \text{Full Evolutions} 				& 1 & 2 & 8 & 64 & 1024 & 32768 & & 35,184,372,088,832  \\ \hline
	  \end{tabular}}
\end{center}
\caption{Counts of distinct representative BFB sequences, evolutions and copy number profiles.}
\label{TableCount}
\end{table}

The resulting counts can be seen in Table 1, where we see the number of sequences grow rapidly with the number of BFBs. We now turn to the enumeration of the number of distinct evolutions for each BFB sequence, where we have thew following result.

\begin{theorem}
If $n$ BFBs take place then the total number of distinct evolutions is given by $2^{\frac{n(n-1)}{2}}$. The total number of evolutions that retain copies of all folds is given by $\prod_{i=1}^n (2^n-1)$. The proportion of evolutions that do not lose information then tends to limit $\prod_{i=1}^n (1-2^{-n})=0.288.$
\label{EvoCount}
\end{theorem}

We can also see these counts in Table \ref{TableCount}, where we see the number of evolutions rising super-exponentially as a function of BFB count, becoming incomputable beyond about eight BFBs.

\begin{figure}[t!]
\centering
\includegraphics[height=125mm,width=125mm]{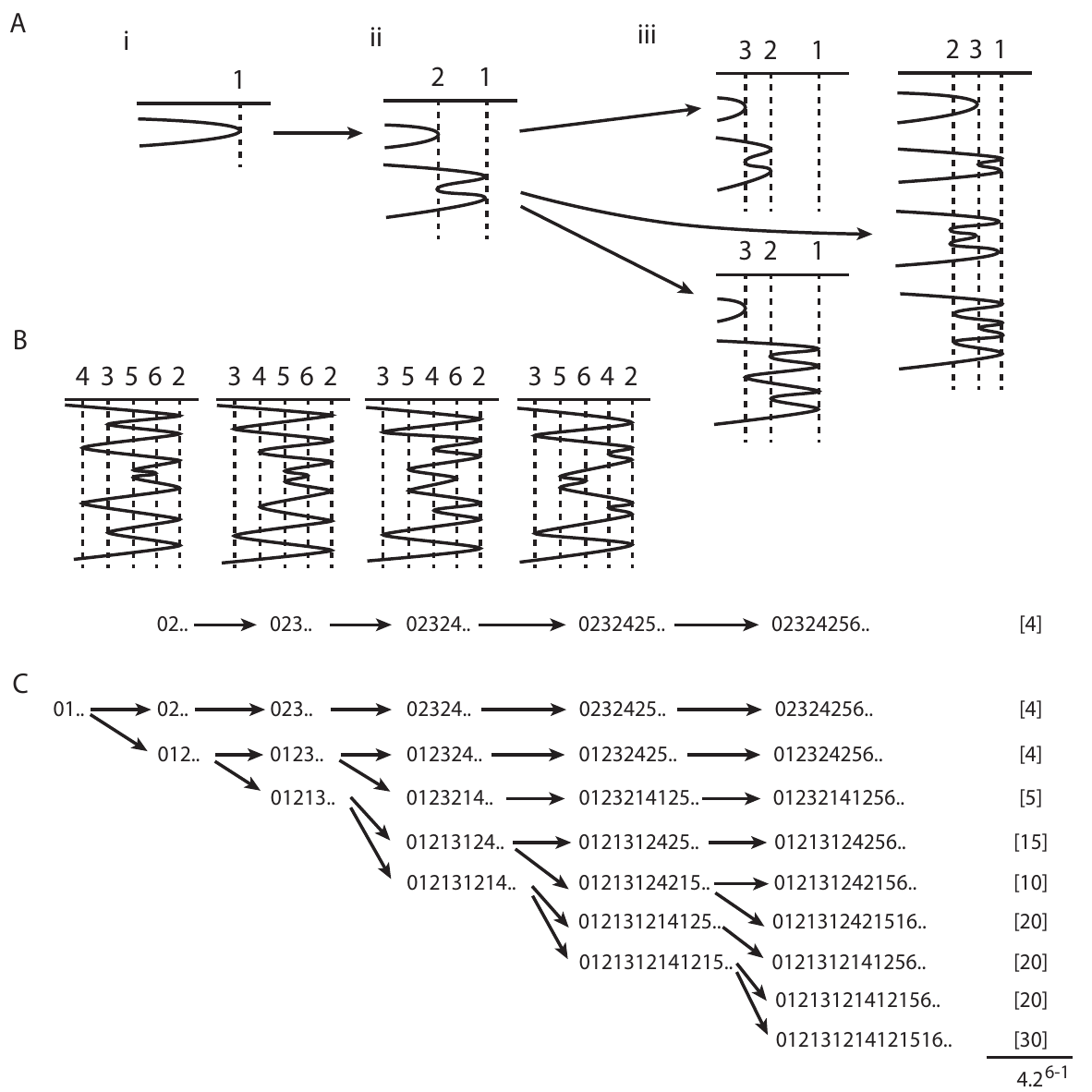}
\caption{Evolutions of BFB cycles. In A we see one structure with one fold (i), two structures with two folds (ii) and $2^{\frac{3(3-1)}{2}}$ structures arising from three folds (iii). In B we have the 5-fold word $023242565242320$ along with $4$ possible structures. In C we see the introduction of a first fold gives rise to $4.2^{5}$ possible structures.}
\label{Doubling}
\end{figure}

The proof of this theorem relies on an appropriate induction. In Figure \ref{Doubling}A we observe the first few evolutions. The initial fold gives rise to one structure (ignoring parity). The structure has two segments that a subsequent fold can form along, resulting in two new structures of Figure \ref{Doubling}Aii. These two structures have two and six segments between them that a new fold can form along, giving eight new structures in Figure \ref{Doubling}Aiii. The number of segments in any structure then determine the number of possible structures we can get with one more fold.

Unfortunately, this does not help us explain Theorem \ref{EvoCount}. Curiously, to prove this we need to reintroduce the \emph{first} fold. In Figure \ref{Doubling}B we see the four possible structures that correspond to the \emph{5-fold} word $W=023242565242320$, along with the sequence of word operations that generate $W$. For each word $W=AnA^{-1}$ in Figure \ref{Doubling}B,C we write $An..$ for brevity. In Figure \ref{Doubling}C we see several possible ways of introducing an initial fold that preserves the order of the other folds in the word. For example, from the word $010$ we can introduce fold $2$ before or after $1$ to give us $020$ or $01210$. The word $020$ then follows the same evolution as Figure \ref{Doubling}B, whereas $01210$ again provides two choices; remove the second $1$ to give $0123210$, or introduce fold $3$ after the second copy of fold $1$ to give $012131210$. When we follow this decision process through all five folds we get nine words. We then calculate the number of orders for each word with Lemma \ref{PosetReduction} and find that we have $2^{5}$ times the original number of orders. To explain this we need to introduce a class of operations on the two-dimensional poset trees introduced above.

We constructed an order tree from the 2d-tree by removing all minor edges. We require the capacity to modify the shape of an order tree with the following operation.
\newline

\begin{figure}[t!]
\centering
\includegraphics[height=40mm,width=60mm]{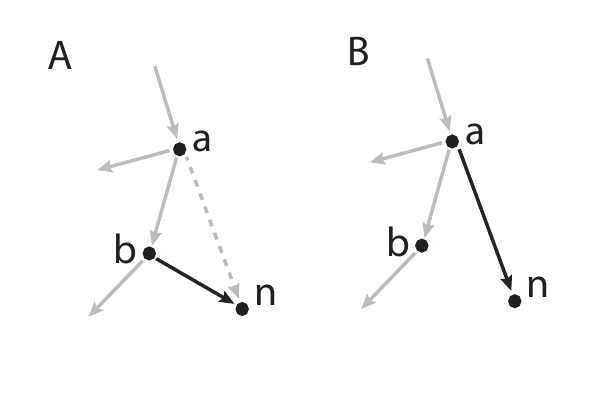}
\caption{Edge Switch operation: Remove the major edge and replace the corresponding minor edge.}
\label{ERO}
\end{figure}

\emph{\bf ES: Edge Switching}

Remove a major edge and replace the corresponding minor edge.
\newline

This move can be seen in figure \ref{ERO}, and effectively moves a branch nearer to the root, and results in a tree structure. This  move has no effect on the other edges or the nodes they are attached to. ES operations thus commute; we can perform the moves in any order and get the same structure.

This is a specific form of the Subtree Prune and Regraft (SPR) operation that has seen application to many other problems in evolution \cite{SempleSteel}.

We are interested in the following sets of ES operations. Let $S(T)$ denote the set of subtrees of a given order tree $T$ that include the root node. Then $T_s$ is the tree obtained from $T$ as follows.
\newline

\emph{\bf SS: Subtree Switching}

\emph{\bf I} Perform an ES operation on any flipped (orange) edge contained in the subtree.

\emph{\bf II} Perform an ES operation on any plain (black) edges adjacent (so not contained) to the subtree.

\emph{\bf III} Leave the remaining edges alone.
\newline

\begin{figure}[t!]
\centering
\includegraphics[height=140mm,width=140mm]{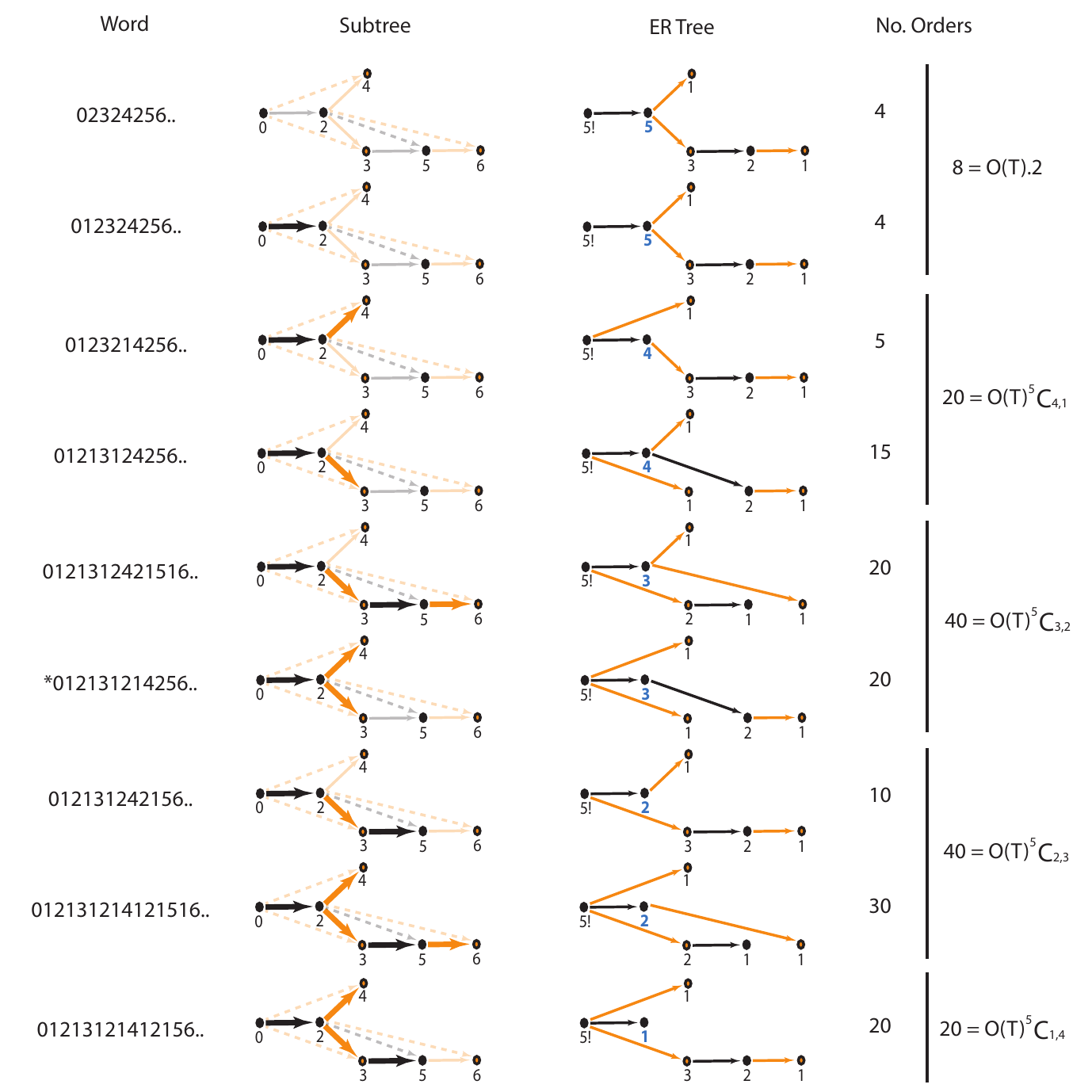}
\caption{Subtree order counts are given for the tree corresponding to word 023242565242320 from Figure \ref{Doubling}. The first column indicates the all nine words arising from introduction of fold $1$. The second column indicates corresponding 2-d tree subtrees in bold. The third column indicates the order tree after implementing SS operations. The fourth column counts the orders.}
\label{TwoPowerGraphs}
\end{figure}

Examples can be seen in Figure \ref{TwoPowerGraphs}. In row $(*)$ we see the subtree with the edges $02$, $23$ and $24$. Edges $23$ and $24$ are flipped, so we replace them with their corresponding minors (3rd column) (SS move {\bf I}). $02$ is untouched. Plain edge $35$ is adjacent to the subtree, so it is also switched (SS move {\bf II}).

We find there is a unique correspondence between the possible introductions of a first fold $1$ and the SS operations.

\begin{lemma}
Let $T$ be the order tree with $n+1$ nodes associated with an n-fold word according to Lemma \ref{PosetReduction}. Let $s$ denote the set of major edges that are directed towards nodes labeled $m$ where we have replaced the recursion $W_m=W_{m-1}(s_m)mW{m-1}(s_m)^{-1}$ with $W_{m-1}(s_m)1m1W_{m-1}(s_m)^{-1}$ in the evolution of the word. Then $s$ is a subtree of $T$ and the corresponding order tree is obtained by implementing SS operations on $s$.
\QED
\label{ERTreeWord}
\end{lemma}

We see examples of this in Figure \ref{TwoPowerGraphs}. The connection between these tree operations and Theorem \ref{EvoCount} can now be introduced.

\begin{theorem}
For any order tree $T$ with $n+1$ nodes, let $S$ denote the set of subtrees with the same root. Then, $\sum_{s \in S}O(T_s)=O(T)2^n$, where $O$ is the order function of Lemma \ref{PosetReduction}.
\label{DoublingResult}
\end{theorem}

This result follows from the following observation, which can be proved inductively.

\begin{lemma}
Let $T$ denote a tree with $n+1$ nodes such that there is exactly one node directly below the root (so $n \ge 1$). For any subtree $s \in S(T)$ we let $b_s$ denote the number of daughter nodes from node $b$, plus one, after implementing SS on the subtree $s$. Then

$\sum_{\{s \in S(T):b_s=r\}}O(T_s)=\left\{
\begin{array}{l l}
{^nC_r}O(T) & r=1,2,...,n-1 \\
2O(T) & r=n \\
\end{array}
\right.$
\QED
\label{TreeTopEdgeLemma}
\end{lemma}

Examples of this can be seen in Figure \ref{TwoPowerGraphs}, where the counts are broken down according to the values of node $b_s$ (in blue).

Summing over the possible values of $r$ then gives us $2^nO(T)$, returning the result of Theorem \ref{DoublingResult}.
\QED

An example of this can be seen in the last column of Figure \ref{TwoPowerGraphs}, where a graph corresponding to \emph{5-fold} word $023242565242320$ and BFB sequence $[1,1,2,2,1]$ with $4$ orders results in $4.2^{5}$ new orders when a first fold is introduced.

We are now in a position to prove our main result and count evolutions. We have seen that any \emph{n-fold} word corresponding to an order tree $T$ has $O(T)$ possible structures. These are associated with $O(T)2^n$ possible n+1-fold structures by the introduction of a new first fold. If we start with the trivial structure and inductively perform these fold introductions, we find that we have $1.2^1.2^2. \hdots .2^{n-1}=2^{1+2+...+n-1}$ possible evolutions. That is, there are $2^\frac{n(n-1)}{2}$ possible evolutions using $n$ folds, as required.

The formula for the reduced sequences is obtained by ignoring the single choice in the introduction of the first fold that loses the first fold (that is, we do not require the first line of Figure \ref{Doubling}C).
\QED

The total number of distinct copy number profiles were also determined for fixed BFB counts, as summarised in Table \ref{TableCount}. Clearly the number of copy number profiles is smaller than the number of evolutions and there may be several evolutions for any given copy number profile.  Furthermore, we can construct an infinite number of BFB sequences with negative values that all reduce to any given reduced sequence. We thus need other methods to help identify the correct evolution for any given copy number profile.

So far we have treated BFB cycles as a discrete process, treating the folded structures as functions of BFB sequences, a sequence space which we have now explored in some detail. However, the BFB process relies on the fold occurring somewhere along the length of the structure. We can thus consider the fold positions in a sequence of BFB structures as a stochastic process, and investigate the implications of this on the BFB structure.


\section{BFBs as a Stochastic Process}

We first consider the stochastic nature of the structures length under the simplest assumption that the fold position is uniformly distributed along the structure. We then consider the likelihood associated with certain BFB representative sequences, showing in particular that structures for some BFB sequences are more likely to occur than others.


\subsection{Length Distributions}

So far we have considered each BFB product as a structure folded with respect to a set of reference positions. We now imagine unfolding the entire structure at each stage.

\begin{figure}[h!]
\centering
  \includegraphics[width=5in,height=2in]{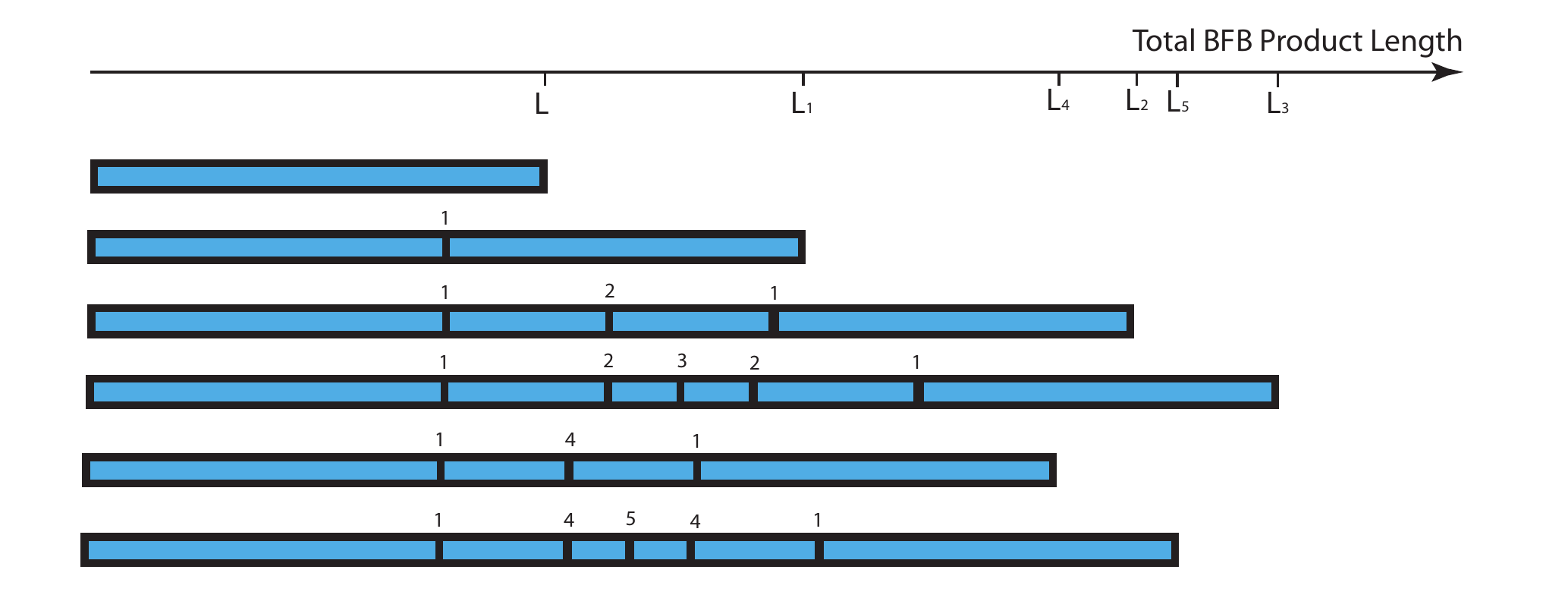}
\caption{The lengths of the unfolded products of a sequence of five BFB cycles}
\label{Lengths}
\end{figure}

Such an example can be seen in Figure \ref{Lengths}. Here we start with a product of length $L_0=L$. From Lemma \ref{Symmetry} we can assume that duplication is on the left side of the position of any BFB fold. The break occurs uniformly along this length, so after duplication, repair and unfolding, the next length $L_1 \in [0,2L_0]$. The first fold in this example is beyond the midpoint of the previous structure, so the resulting length increases, as it does for the next two BFBs. However, the fourth fold occurs in the first half of the previous structure, reducing the length and removing the second and third folds before the final BFB again extends the structure.

We then see that the length $L_n$ is a stochastic Markovian process with conditionally uniform distribution $(L_n|L_{n-1}) \sim U([0,2L_{n-1}])$. The general length distribution $P(L_n)$ can then be derived, giving the following result.

\begin{theorem}
If $L_0=L$ is the initial length of the chord, then the length $L_n$ after the $n^{th}$ BFB has distribution

$P(L_n)=\left\{
\begin{array}{l l}
\frac{1}{2^n\Gamma(n) L}log^{n-1}\left(\frac{2^{n+1}L}{L_n}\right) & L_n \leq 2^nL \\
0 & L_n>2^nL \\
\end{array}
\right.$

with mean value $L$ and standard deviation $L\sqrt{\left( \frac{4}{3} \right)^n-1}$.
\label{LengthTheorem}
\QED
\end{theorem}

Thus we find that although the lengths average value does not change, it is increasingly variable. We also see that the shortest distance $\frac{L_m}{2}$ of any copy of the $m^{\emph{th}}$ fold from the ends is preserved. The first fold encountered is always a distance $\frac{L_1}{2}$ from either end, for example. All copies can be lost however, and we have seen in Figure \ref{Lengths} that as the BFB process continues,  BFB events that shorten the structure can delete all copies of folds from some previous BFB events. We can characterize these properties as follows.

\begin{theorem}
The original distance $\frac{L_m}{2}$ of the $m^{\emph{th}}$ BFB fold from the end of the structure is the shortest distance of any subsequent copy of that fold to either end. If $L_n<L_m$  and $n>m$ then all copies of the $m^{\emph{th}}$ BFB fold are permanently excised from the BFB product. Thus if the $m^{\emph{th}}$ BFB fold is to avoid extinction through a series of $N$ BFBs then $L_m<min_{\{n>m\}}L_n$. Subsequently, if we have a series $L_1,L_2,...,L_N$ of BFB lengths, the only BFB folds that survive will be a subset with increasing length, in the same order that they occurred.
\label{SurvivalCriterion}
\QED
\end{theorem}

This raises two issues. Firstly, if we observe a sequence $L_1<L_2<...<L_N$ of BFB lengths in a final structure we would like to know how many folds from other BFB events have been completely excised from the genome in the process. Secondly, we know that the smallest length $L_1$ is the earliest remaining BFB. The fold at position $\frac{L_1}{2}$ is thus the first encountered as we traverse the structure. This also gives the position of the outermost fold relative to the reference. For example, in Figure \ref{Posets}Ai, the first fold, at position $x_1$, is furthest from the ends of the structure, relative to the reference positions. The first fold thus measures the size of the amplicon.

A better understanding of the order statistics of the length sequence $L_n$ will help our understanding of both the scale of deleted BFB folds, and the size of amplicons, where we have the following result.

\begin{theorem}
The probability density $M_{k,N}(x,L)$ that the $k^{\emph{th}}$ BFB of a series $L_1,L_2,...,L_N$ is the minimum with length $x$ is given by

$M_{k,N}(x,L)=\frac{1}{2^{k}L}W_k(x,L)(1-\sum_{i=1}^{N-k}\frac{1}{2^i}W_i(x,x))$

where $L$ is the original length and,

$W_k(x,y)=\int_x^{2y}\int_x^{2z_1}...\int_x^{2z_{k-1}}\frac{1}{z_1...z_k}dz_k...dz_1 = \sum_{j=0}^{k}a_{j+1}^k(x) \log^j(2^ky)$,

$a^k$ is the $k+1$ length vector $\prod_{r=1}^kB_r$, and $B_r$ is the $(r+1) \times r$ matrix

$
\left( 
\begin{array}{c c c c c}
-\log(2^{r-1}x) & -\frac{1}{2}\log^2(2^{r-1}x) & \cdots & -\frac{1}{r}\log^r(2^{r-1}x) \\
1 & 0 & \cdots & 0 \\
0 & \frac{1}{2} & \cdots & 0\\
\vdots & \vdots & \ddots & \vdots \\
0 & 0 & \cdots & \frac{1}{r}\\
 \end{array} 
\right)
$
\label{OrderTheorem}
\QED
\end{theorem}

We can use this to get the distribution of both the minimum length and its occurrence in the BFB sequence, as indicated in Corollary \ref{LengthCor}i-iii below.

The result also enables us to get the distribution of the amplicon size, that is, the position $L_{amp}$ of the outermost fold relative to the reference, $\frac{1}{2}min_{k \le N}\{L_k\}$, as summarized in Corollary \ref{LengthCor}iv.
 
We next consider an observed sequence of BFB folds with corresponding lengths $L_1<L_2<L_3<...<L_n$ and estimate how many BFBs were likely to have been deleted in this process. Specifically, if we have a sequence of BFBs with lengths $l_{1,1},l_{1,2},...,l_{1,d_1},$ $L_1,l_{2,1},l_{2,2},...,l_{2,d_2},L_2,...,L_n$
such that $L_{i-1} \le L_i \leq {l_{i,1},...,l_{i,d_i}}$, then by Theorem \ref{SurvivalCriterion} the BFB with length $L_i$ deletes the $d_i$ earlier BFB folds with longer lengths ${l_{i,1},...,l_{i,d_i}}$ to leave the events $L_{i-1},L_i$. We can use Theorem \ref{OrderTheorem} to estimate the scale of loss, $d_i$. If we have a BFB of length $L_{i-1}$, which is followed by the sequence $l_{i,1},...,l_{i,d_i} \ge L_{i-1}$, then we require $l_{i,1},...,l_{i,d_i}>L_i$, given that we start with length $L_{i-1}$. That is $Pr(l_{i,1},...,l_{i,d_i} \ge L_i|L_{i-1})  =\frac{1}{2^dL_{i-1}}\int_{L_i}^{2L_{i-1}}\int_{L_i}^{2l_1}...\int_{L_i}^{2l_{d-1}}\frac{1}{l_1....l_{d-1}}dl_d...dl_1$. This can be calculated in much the same way as Theorem \ref{OrderTheorem} (see Appendix). A Bayesian inversion then allows us to estimate the distribution of $d_i$, given in Corollary \ref{LengthCor}v. In summary we have the following.

\begin{corollary}
(i) The probability that the $k^{th}$ of $N$ BFBs is the one with the minimum length $L_{min}$ is given by $\frac{M_{k,N}(L_{min},L)}{\sum_{k=1}^NM_{k,N}(L_{min},L)}$.

(ii) The probability density function of the minimum BFB length in a sequence of $N$ BFBs is given by $\sum_{k=1}^NM_{k,N}(L_{min},L)$.

(iii) The distribution of the number $N$ of BFBs for a given minimum length $L_{min}$ is then given by $Pr(N=n|L_{min})=\frac{\sum_{k=1}^nM_{k,n}(L_{min},L)}{\sum_{n=1}^{\infty}\sum_{k=1}^nM_{k,n}(L_{min},L)}$.

(iv) The amplicon size, $L_{amp}$, given a sequence of $N$ BFB cycles has distribution $2\sum_{k=1}^NM_{k,N}(2L_{amp},L)$.

(v) If we observe two BFBs with consecutive lengths $L_{i-1}<L_i$ the number of BFBs occurring between them that are deleted by the $i^{th}$ BFB, $D_i$, is, 
$Pr(D_i=d|L_{i-1}<L_i)=\frac{I_d}{\sum_{d=1}^{\infty}I_d}$, where $I_d=1-\frac{L_i}{2L_{i-1}}\sum_{k=0}^{d-1}\frac{1}{2^k}W_k(L_i,L_{i-1})$ and $W_0(L_i,L_{i-1})=1$.

\label{LengthCor}
\QED
\end{corollary}

\begin{figure}[t!]
\centering
\includegraphics[height=100mm,width=115mm]{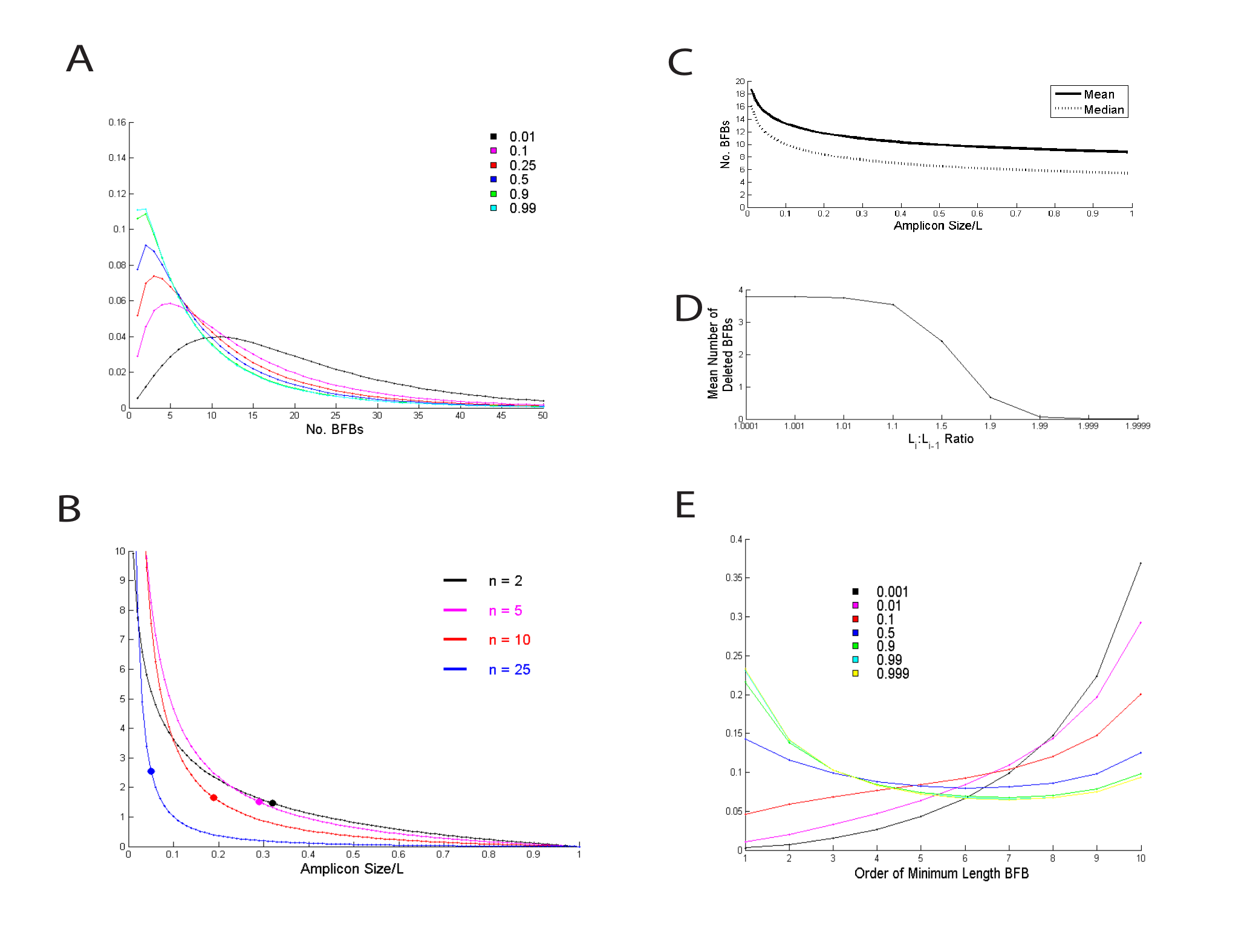}
\caption{BFB Distributions. A) The distribution of BFB counts for a range of amplicon sizes (as a proportion of $L$). B) The distribution of the amplicon size for a range of BFB counts. The mean positions are located at the circles. C) The mean and median number of BFBs as a function of amplicon size. D) The expected number of deleted BFB events as a function of the ratio $L_i:L_{i-1}$. E) The distribution of the minimum length BFB for ten BFBs and a range of amplicon sizes. }
\label{Stats}
\end{figure}
 
Some of these distributions are plotted in Figure \ref{Stats}, where we see the trend that the outermost fold of the BFB, that is, the size of the amplicon, decreases as the number of BFB events increases.

This can be intuited as follows. If a BFB product has current minimum length $L_{min}$, then there is a chance the next fold will be smaller than $\frac{L_{min}}{2}$, deleting all previous folds, and reducing the position of the outermost fold. That is, the BFB process will result in atrophy of the amplicon size.

We also know from Theorem \ref{LengthTheorem} that the average length of the structure does not change. This means that, on average, the same amount of DNA is present in a diminishing region of the reference genome, resulting in the localized high copy number structures typical of amplicons, such as Figure \ref{Amplicon}C.


\subsection{Fold Structure Likelihoods}

We have seen different BFB structures arising due to different BFB sequences. It is thus natural to investigate the likelihood of a particular BFB sequence occurring. We extend the stochastic process approach above to elucidate this problem.

Suppose we are interested in the likelihood of observing BFB sequence such as $r=[1,1,2]$, the fourth structure in Figure \ref{Intro}. We can build the likelihood inductively. The first fold $x_1$ is uniformly distributed across the original structure of length $L$, so we have $Pr(x_1|r_1)=\frac{1}{L}$. The second fold occurs at position $x_2$ on the second segment ($r_2=1$) and so satisfies the inequality $x_0<x_2<x_1$. It is uniformly distributed along this segment, so $Pr(x_2|x_1,r_1,r_2)=\frac{1}{x_1-x_0}$. Now $Pr(r_3=2|x_1,x_2,r_1,r_2)$ is the chance of hitting the last of the four segments of the structure corresponding to $[r_1,r_2]=[1,1]$. This segment has length $x_1-x_0$ and the total length of the structure is $2(x_1-x_0)+2(x_1-x_2)=4x_1-2x_2-2x_0$. If we suppose $x_0=0$ and $L=1$ for simplicity then we get probability $\frac{x_1}{4x_1-2x_2}$. We can then can put this information together to get the probability of getting BFB sequence $[1,1,2]$ conditional upon $[1,1]$ as follows:

\begin{align}
\begin{split}
\nonumber
P([1,1,2]|[1,1]) & =\int_{0<x_2<x_1<1}P([1,1,2],x_1,x_2|[1,1])dx_1dx_2\\
& =\int_{0<x_2<x_1<1}P([1,1,2]|x_1,x_2,[1,1])P(x_2|x_1,[1,1])P(x_1|[1,1])dx_1dx_2\\
& =\int_{0<x_2<x_1<1}P(r_3=2|x_1,x_2,r_1,r_2)P(x_2|x_1,r_1,r_2)P(x_1|r_1)dx_1dx_2\\
& =\int_{0<x_2<x_1<1}\frac{x_1}{4x_1-2x_2}.\frac{1}{x_1}.\frac{1}{1}dx_1dx_2=\frac{1}{2}\log 2\\
\end{split}
\end{align}

\begin{figure}[ht!]
\centering
  \includegraphics[width=80mm,height=100mm]{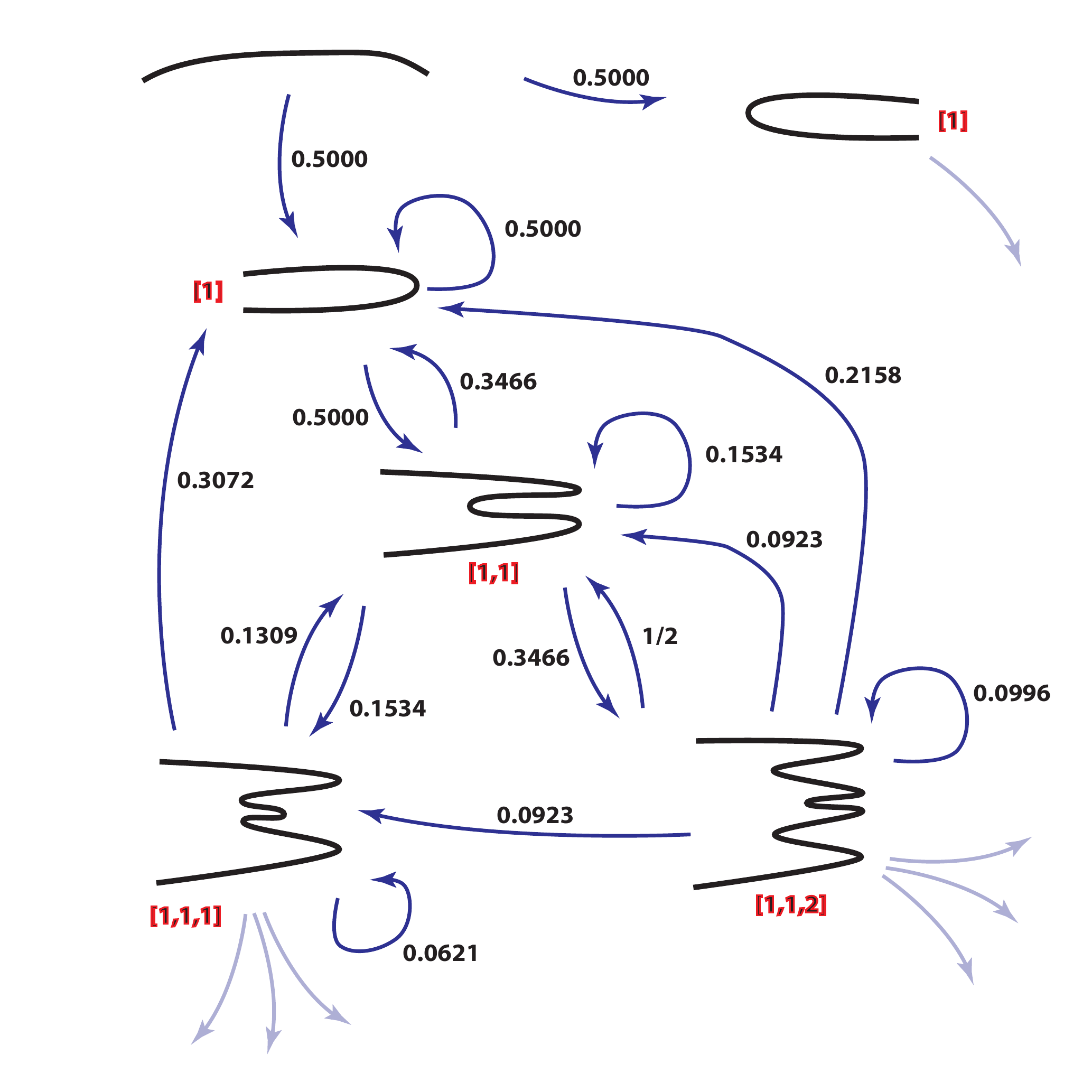}
\caption{Probabilities of fold events in BFB space. The black chords indicate the structure for each BFB sequence (indicated in red). Numbers alongside arrows indicate the probability of taking that step.}
\label{Tree}
\end{figure}

This process can be applied in general which we summarized below.

\begin{lemma}
\label{Likelihood}
The likelihood of seeing reference positions ${x_n}$ for a given BFB sequence ${r_n}$ is given by,
$P(x_1,x_2,...,x_n|r_1,r_2,...,r_n)=\prod_{i=1}^{n} \{ \frac{1}{x_{i_{max}}-x_{i_{min}}}\}$,
where $x_{i_{min}}<x_i<x_{i_{max}}$ are the inequalities of the poset for ${r_n}$ given by Lemma \ref{PosetDefine}.

The conditional probability of next BFB sequence element $r_n$ is $Pr(r_n|x_1,...,x_{n-1},$ $r_1,...,r_{n-1})=\frac{x_{n_{max}}-x_{n_{min}}}{L_{n-1}(x_1,...,x_{n-1})}$, where $L_{n-1}$ is the length after $n-1$ BFB cycles, a linear homogeneous function of $x_1,...,x_{n-1}$. We then find 

\begin{align}
\begin{split}
\nonumber
Pr(r_n|r_1,...,r_{n-1}) & =\int_{\Delta}Pr(r_n,x_1,...,x_{n-1}|r_1,...,r_{n-1})d{\bf x} \\
& =\int_{\Delta}\frac{x_{n_{max}}-x_{n_{min}}}{L_{n-1}(x_1,...,x_{n-1})}.\prod_{i=1}^{n} \{ \frac{1}{x_{i_{max}}-x_{i_{min}}}\}d{\bf x},
\end{split}
\end{align}

where $\Delta$ is the region defined by inequalities $x_{i_{min}}<x_i<x_{i_{max}}$.
\QED
\end{lemma}

The probabilities for the first few BFB sequences can be seen in Figure \ref{Tree}. These integrals rapidly become intractable and numerical methods are required. The simplest method is to randomly generate $x_1,x_2,..$ according to the conditional uniform distributions and average the simulated probabilities $P(r_n|x_1,x_2,...,x_{n-1},r_1,...,r_{n-1})$.

Note that the first BFB has a 0.5 probability of having positive or negative parity, this is not encapsulated by this formula. Note also that these probabilities are not for reduced BFB sequences. For example, although BFB sequences $[1,1,1]$ and $[1,1,2,-1]$ reduce to the same structure, they take different paths through the evolutionary graph of Figure \ref{Tree} and have different likelihoods of occurrence. Multiplying the edge probabilities, we find the probabilities of arising are 0.038  and 0.008, respectively.


\section{Applications to Amplicons in Cancer Genomes}

We now put the structure we have described into context with some real data.


\subsection{Inference of BFB evolution}

For a given amplicon such as Figure \ref{Amplicon}C, identifying the underlying BFB sequence is desirable because it provides an explanatory evolution of events. This would allow us to obtain the order that the folds occurred in this process, and hence construct the fold structure that generates the amplicon.

We would like to use the machinery developed above. There are some extra difficulties, however. Although the experimental signal (read depth) is a linear function of copy number, in many cases it can be difficult to ascertain the actual integer copy number of each segmented region with methods such as \cite{Greenman2}, especially when the regions are small. However, the signals across the amplicon can often be ranked, and the likely number of BFBs identified, meaning we can filter the set of possible evolutions to a more manageable set.

\begin{table}
\begin{center}
 \scalebox{0.8}{%
  \begin{tabular}{| c || c | c | c | c | c | c |}
    \hline
    \text{Rank} & BFB Sequence & Copy Number Profile \textdagger & Order & $\log Pr({\bf x}|{\bf r})$ & $\log Pr({\bf z}|{\bf c})$ & Log-Likelihood \\ \hline
    \text 1 & [1,1,2,2,3] & [16,12,14,6,2] & [1,5,4,2,3] & -81.674 & -921.1 & -1002.774 \\ \hline
    \text 2 & [1,1,2,4,1] & [16,12,14,6,2] & [1,4,2,5,3] & -82.264 & -921.1 & -1003.364 \\ \hline
    \text 3 & [1,1,2,4,5] & [24,20,22,10,2] & [1,4,5,2,3] & -82.264 & -2008.9 & -2091.164 \\ \hline
    \text 4 & [1,1,2,2,1] & [12,8,10,6,2] & [1,5,2,4,3] & -81.674 & -2056.7 & -2138.374 \\ \hline
    \text 5 & [1,1,2,2,5] & [20,12,14,10,2] & [1,5,2,4,3] & -82.038 & -2076.4 & -2158.438 \\ \hline
  \end{tabular}}
\end{center}
\caption{Top five likelihoods for evolution of Figure \ref{Amplicon}C. \textdagger Copy numbers only include  regions II - VI for chromosome undergoing BFB process, other chromosomes are ignored.}
\label{TableLikelihood}
\end{table}

For example, consider the amplicon of Figure \ref{Amplicon}C. We have six segmented regions $I$ to $VI$, separated by five breakpoints. For four of these we used next generation sequencing to find rearrangement positions \cite{Greenman1} and found discordantly mapping reads consistent with BFB events. Although no aberrantly mapping reads could be found at the junction between regions $I$ and $II$, this was likely due to mapping difficulties and the data are indicative of a structure formed by BFB cycles. The rightmost region, $VI$, has a higher signal than the leftmost, $I$, suggesting a BFB with right parity ($p=1$); region $I$ is not part of the BFB structure. This gives us five segmented regions, and so five folds to explain. The fold positions are labeled $x_i, i=1,...,5$. We select the most likely evolution as follows.

The mean (sequence depth) signals for the six regions were ${\bf z}=[154.7,519.8,398.2,$ $465.2,305.5,186.3]$. This is based upon ${\bf m}=[5578,6716,3969,2536,8768,5366]$ measurements (bins containing reads) in each region. We also measure the standard deviations in each region ${\bf \sigma}=[31.4,80.1,63.1,67.8,69.5,45.1]$. Then we assume that for any given evolution with BFB sequence ${\bf r} = [r_1,...,r_n]$ and copy number profile ${\bf cn} = [c_1,...,c_n]$, we have normal distribution $(z_i|c_i) \sim N(\alpha + \beta c_i,\frac{\sigma_i^2}{m_i})$, where $\alpha$ and $\beta$ are parameters that represent the linear relationship between signal and integer copy number. If we have good information on this relationship $\alpha$ and $\beta$ can be stated, otherwise we treat them as unknown parameters. We then construct a likelihood $Pr({\bf z},{\bf x}|{\bf r,c})=Pr({\bf z}|{\bf c})Pr({\bf x}|{\bf r})$ with the help of Lemma \ref{Likelihood}. This is maximized over $\alpha$ and $\beta$ and the likelihood recorded.

For Figure \ref{Amplicon}C we found 24 evolutions produced copy number profiles with the same rank as $z_i$, arising from 9 distinct BFB sequences. The top five are listed in Table \ref{TableLikelihood}. The maximum likelihood solution suggests the actual copy numbers are $[16,12,14,6,2]$, corresponding to the BFB sequence $[1,1,2,2,3]$. The resultant genome is given in Figure \ref{Amplicon}D.

We note that even with all this information we cannot necessarily guarantee a strongly identified best fit. For example, we see the top two ranked evolutions produce the same copy number profile and it is only the likelihood $Pr({\bf x}|{\bf r})$ that weakly distinguishes these two cases. There is the possibility of utilizing additional information in the form of single nucleotide mutations within the amplicon. This is an approach the has been applied in a more general context previously \cite{Greenman1}, which can also be used to time rearrangement events relative to the nucleotide mutation process. However, amplicons are often quite narrow and may not contain sufficient mutations to give much statistical power to further differentiate evolutions and was not explored further.

\begin{figure}[ht!]
\centering
  \includegraphics[width=80mm,height=60mm]{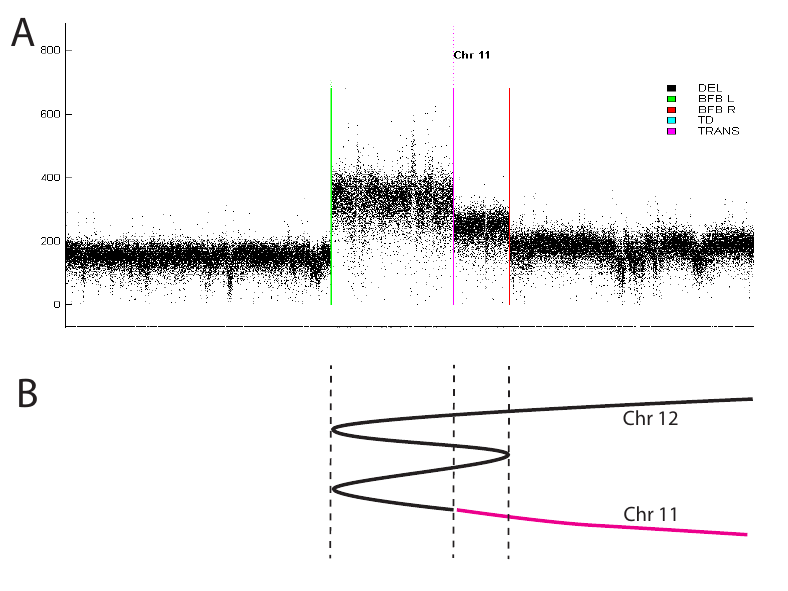}
\caption{A BFB cluster. In A we see the amplicon, the two outer breakpoints are BFBs, the middle one a translocaiton to chromosome 11. B indicates the genomic structure.}
\label{Translocation}
\end{figure}

The portrait of BFB cycles sketched in Figure \ref{Amplicon}A would appear to continue indefinitely. However, this process will stop if we only have one centromere after DNA repair. This may be because a somatic telomere forms on one of the broken ends, but can also be because one of the exposed ends is attached to a different exposed end, such as another chromosome. This pattern can be observed in the data where, in Figure \ref{Translocation} for example, we see a region with three breakpoints in chromosome 12, two of which are assoicated with BFB folds, and the middle one associated with a translocation to chromsome 11. This was likely to be the last step, terminating the breakage fusion bridge process.
  
\subsection{Clonality of BFB Process Under Selection}

Our models have so far assumed that the BFB mutation process is randomly sampled and under no forces of selection. This is unlikely to be the case in general and the selection of any cells that have growth advantage are likely to emerge in cancer samples. We have seen in Figure 1A that the BFB sequence arises due to spindles attached to the dicentromeric chromosome, dissecting the chromosome during cell division. This preserves the total amount of doubled DNA in both daughter cells. For a given break, we find the DNA is duplicated one one side of the break in one daughter cell, and on the other side in the other daughter cell. In Figure \ref{Amplicon}A, for example, one daughter cell contains two yellow and four red genes, the other daughter cell two yellow genes, but the total number of red and yellow genes across both progeny is conserved at four. We then find that if the parent cell has BFB sequence $[r_1,r_2,...,r_{n-1}]$ with $s_{n-1}=\sum_{i=1}^{n-1}r_i$, and one daughter cell has sequence $[r_1,r_2,...,r_{n-1},r]$, then the other daughter cell has sequence $[r_1,r_2,...,r_{n-1},1-r]$. Given that we must have $-s_{n-1} < r \le s_{n-1}$, one of $r$ and $1-r$ must be non-positive.

If this process proceeds over $n$ cycles, and so $n$ cell divisions, producing $2^n$ cells, we are left to conclude that one of the cell lineages will consist entirely of positive terms and hence lose no BFB folds. This lineage is always gaining DNA by Theorem \ref{BFB_Seq} and will be a good candidate to contain multiple copies of genes that may be advantageous to cancer. This cell may then emerge as a dominant clone, resulting in subsequent rapid amplicon development.

We can also argue this from a different perspective. Note that the distribution of the amplicon size in Figure \ref{Stats}B has a mean value that moves toward the origin as the number of BFBs increases. If a gene is, for the sake of argument, half way between centromere and telomere, and five BFB cycles take place, one can integrate this distribution up to that point to conclude that there is approximately a 95 \% chance that the outermost fold is before the target gene and therefore only one copy of the gene is present (on the other allele) in the cell. Initially each cell has two copies of a gene. After 5 cell divisions there will be 32 cells and 64 copies of the gene target distributed amongst them. This implies that many copies of those genes are likely to be contained in one or two of those cells. Thus we find that it only takes relatively few BFB cycles to generate a cell containing multiple copies of a gene. If the gene is an oncogene, this cell then becomes a good target for selection and subsequent clonal expansion, producing the types of amplicons observed in cancer.

Selection thus plays a fundamental role in the evolution of these structures and a fuller investigation of selection acting across a growing set of cells undergoing a BFB process is warranted.


\section{Conclusions}

We have highlighted some of the genomic complexities that arise from the BFB process that underlies the copy number profile of many amplicons observed in cancer. Although not every copy number profile can arise from a BFB process, the number of different BFB evolutions rises spectacularly quickly with the number of BFB cycles. Furthermore, a single copy number profile may be possible from more than one BFB evolution, complicating the inference of the correct evolution. For such degenerate cases, use of additional in-silico methods such as \cite{Greenman1}, or experimental methods such as Fluorescent In Situ Hybridisation (FISH), will be necessary to help identify the actual chromosomal structure and underlying process.

This work provides some understanding to the evolution of amplicons. However, amplicons can arise from other processes such as tandem duplication \cite{McBride} or double minutes \cite{Raphael1}, for example, and amplicon evolution in general will be somewhat more complicated, possibly involving combinations of these processes, as well as other unexplored mechanisms.

This analysis also assumes that the data arise from a single dominant clone, which is not always the case \cite{Nik-Zainal1, Nik-Zainal2}. All of these other factors will have to be taken into account if we are to unravel more general evolutions of amplicons. However, the work presented is one step in that direction.


\nonumber\section{Acknowledgements}
We acknowledge insightfull discussions with Katharina Huber.

\appendix
\section{Appendix - Proofs}

\begin{proof}
{\bf Lemma \ref{Symmetry}}. Suppose we have a palindromic region word of the form $XYY^{-1}X^{-1}$ such that the next BFB fold occurs between $X$ and $Y$. If we duplicate the left (resp. right) side of the fold, we get word $XX^{-1}$ (resp. $XYY^{-1}YY^{-1}X^{-1}$). We can similarly have a breakpoint at the symmetrically opposite position between $Y^{-1}$ and $X^{-1}$. Duplication of the left (resp. right) side of the fold then produces words $XYY^{-1}YY^{-1}X^{-1}$ (resp. $XX^{-1}$). These events cannot be distinguished as they have identical products.
\end{proof}

\begin{proof}
{\bf Theorem \ref{Algorithm}}. The final product of a BFB sequence must have a fold word of the form $XnX^{-1}$ and so we can select the middle symbol, which must be unique as it is the last fold to form. Undoing this fold gives us the word $X$ to consider. Now $X$ is a product of a BFB, but fold $n$ may have truncated some sequence $Z$. $X$ must thus have the form $ZYmY^{-1}$ for some subwords $Y$ and $Z$ (where $-1$ power indicates symbols in reversed order). Now in any word generated by a BFB process, if we have two consecutive occurrences of a symbol $m_1$ then there must have been a BFB with fold $m_2$ that duplicated $m_1$ with fold $m_2$ between the two copies. Thus fold $m_2$ occurred later than $m_1$ and we have a word of the form $...m_1...m_2...m_1...$ . Note that the leftmost position of $m_1$ does not change position in the word when fold number $m_2$ is incorporated. The leftmost $m_2$ is to the right of the leftmost $m_1$ and we see that the first occurrences of fold symbols reflect their evolutionary order. If there is more than one symbol $m_2$ we can repeat the procedure, forming series $m_1,m_2,...$ until we find a symbol $m_n$ occurring once. This must exist because each fold symbol $m_i$ is located further into the word than $m_{i-1}$, and the word is finite in length. There may be more than one symbol occurring once (resulting in a word of the form $Xm_nm_{n+1}...m_{n+u}$ for unique symbols $m_n,...,m_{n+u}$). The rightmost symbol $m_{n+u}$ must then be the latest event that we undo in STEP 2. Because the word is reduced in size at each step we either obtain a valid BFB evolution or the algorithm fails and the word is not a viable representation of a BFB process.
\end{proof}

\begin{proof}
{\bf Theorem \ref{BFB_Seq}}. We show that $s_n$ counts positively labeled segments of the $n^{\emph{th}}$ BFB structure inductively. This is true for $n=1$ because we start with $s_1=r_1=1$ segment with label $1$. Assume true for $n=k$ so that we have $s_k>0$ positive values labeling the segments after the midpoint, and so $2s_k$ labels in total. Then we can select label $r_{k+1}$ for the next BFB with $-s_k<r_{k+1} \le s_k$. This means that if $r_{k+1}>0$ we duplicate $s_k$ non-positive and $r_{k+1}$ positive labels, producing $s_k+r_{k+1}=s_{k+1}>0$ new positive labels, as required. if $r_{k+1}<0$ then $s_k-(-r_{k+1})$ counts the number of (negative) labels that are duplicated, again producing $s_k+r_{k+1}=s_{k+1}>0$ new positive labels. Thus $s_n$ always counts the number of positively labeled segments.

We then find that the structure formed by the $n^{\emph{th}}$ BFB cycle has $2s_n$ segments, $s_n$ with positive labels and $s_n$ with non-positive labels, separated by the structures midpoint, which is the position of the fold formed by the $n^{\emph{th}}$ BFB cycle. Subsequently, $s_n$ counts the number of segments we traverse through the structure until we encounter this fold, as required.

Now, the $(k-1)^{\emph{th}}$ and $k^{\emph{th}}$ BFB folds occur on the $s_{k-1}^{\emph{th}}$ and $s_k^{\emph{th}}$ segment. Now if $r_k=s_k-s_{k-1}<0$, fold $k-1$ is positioned further into the structure than fold $k$, and so is deleted. Then the $(k-1)^{\emph{th}}$ fold can be removed and cumulative sequence $[...,s_{k-2},s_{k-1},s_k,...]$ can be replaced with $[...,s_{k-2},s_k,...]$. Taking differences between consecutive terms, BFB sequence $[...,r_{k-2},r_{k-1},r_k,...]$ then becomes $[...,r_{k-2},s_k-s_{k-2},...]=[...,r_{k-2},(s_k-s_{k-1})+(s_{k-1}-s_{k-2}),...]=[...,r_{k-2},r_{k-1}+r_k,...]$, one term has been absorbed and one BFB has been removed.

We have seen that the $k^{\emph{th}}$ BFB fold is located on the $s_k^{\emph{th}}$ segment. The first fold of this structure points in direction $-p$, and the direction alternates with segments. We thus find that the $k^{\emph{th}}$ fold points in direction $p(-1)^{s_k}$, as required.
\end{proof}

\begin{proof}
{\bf Lemma \ref{PosetDefine}}. The fold word symbols indicate the order of folds occurring in the structure. From Theorem \ref{BFB_Seq}, if $r_n$ is the next value in the BFB sequence, the segment this fold occurs on is $s_n$ from the end. This stretches between values $x_{W_{n-1}(s_n)}$ and $x_{W_{n-1}(s_n+1)}$. The value $s_n$ counts the segments from the start of the structure, which alternate in direction, so by Therorem \ref{BFB_Seq}, $d_n$ indicates which of $x_{W_{n-1}(s_n)},x_{W_{n-1}(s_n+1)}$ is larger, giving the inequalities specified.
\end{proof}

\begin{proof}
{\bf Lemma \ref{MinorParent}}. We have two cases to consider. For case I, if node $n$ formed from $a$ and $b(>a)$ is plain, then we have word $..ab..$ becoming $..ana..$. When we then connect a new node $n'$ we have $..ana..$ becoming either $..an'a..$ or $..ann'na..$. In either case we construct a major edge from $n$ to $n'$, and a minor edge from $a$ to $n'$. For case II, if node $n$ formed from $a$ and $b$ is flipped, then we have word $..ba..$ becoming $..bnb..$. When we then connect a new node $n'$ we have $..bnb..$ becoming either $..bn'b..$ or $..bnn'nb..$. In either case we construct a major edge from $n$ to $n'$, and a minor edge from $b$ to $n'$.
\end{proof}

\begin{proof}
{\bf Lemma \ref{BFB_Counts}}. Here we let $v_{n,m}$ represent the number of reduced BFB sequences such that $s_n=\sum_{k=1}^nr_k=m$. $v_{1,n}$ is thus zero apart from the first unit entry. By the definition in Theorem \ref{BFB_Seq} of reduced BFB sequences, each sequence $[r_1,r_2,...,r_n]$ can have a subsequent positive entry $r_{n+1} \epsilon \{1,2,...,m\}$. Thus, conversely, a sequence $[r_1,r_2,...,r_{n+1}]$ of length $n+1$ and total $m$ must contain the sub-sequence $[r_1,r_2,...,r_n]$ with a total ranging from $\lfloor\frac{m+1}{2}\rfloor$ to $m-1$. This gives us the relation $v_{n+1,m}=\sum_{k=\lfloor\frac{m+1}{2}\rfloor}^{m-1}v_{n,k}$. Summing $v_{m,n}$ over second index $m$ then counts the number of representative sequences of length $n$.

Similarly, we let $w_{n,m}$ represent the number of full representative sequences $[r_1,r_2,...,r_n]$ such that $s_n=\sum_{k=1}^nr_k=m$. $w_{1,n}$ is thus zero apart from the first unit entry. By the definition in Theorem \ref{BFB_Seq} of full BFB sequences, each such sequence can then have a subsequent entry $r_{n+1} \epsilon \{-(m-1),-(m-2),...,0,1,2,...,m\}$. Thus, conversely, a sequence $r_1,r_2,...,r_{n+1}$ of length $n+1$ and total $m$ must contain the sub-sequence $r_1,r_2,...,r_n$ with a total ranging from $\lfloor\frac{m+1}{2}\rfloor$ to $\infty$. This gives us the relation $w_{n+1,m}=\sum_{k=\lfloor\frac{m+1}{2}\rfloor}^{\infty}w_{n,k}$. Summing $w_{n,m}$ over second index $m$ then counts the number of full representative sequences of length $n$.
\end{proof}

\begin{proof}
{\bf Lemma \ref{ERTreeWord}} We have three cases to consider.

For SS move I, suppose we have word $..ba..$ ($b>a$) becoming $..bmb..$. Then we have a minor flipped edge from $a$ to $m$ and major flipped edge from $b$ to $m$. To introduce first fold $1$ adjacent to $m$, we must start with a word of the form $..b1a..$. Node $b$ is adjacent to fold number $1$ and is part of subtree $s$. We then find that $..b1a..$ becomes either $..b1m1b..$ or $..bmb..$. From Lemma \ref{MinorParent}, in the former case we find node $m$ has a major edge connected to $a$, that is we have switched the major for minor edge and performed an ES operation. In the latter case we get the same result as before and no changes are made to major/minor status. Because $m$ is not adjacent to fold number $1$, edge $bm$ is not part of the subtree.

For SS move II, suppose we have word $..ab..$ ($b>a$) becoming $..ama..$. Then we have a minor plain edge from $a$ to $m$ and major plain edge from $b$ to $m$. To introduce first fold $1$ adjacent to $m$, we must start with a word of the form $..a1b..$. Node $b$ is adjacent to fold number $1$ and is part of the subtree. We then find that $..a1b..$ becomes either $..a1m1a..$ or $..ama..$. From Lemma \ref{MinorParent}, in the former case we find node $m$ has a major edge connected to $b$, that is we have the same major edge. Because fold $1$ is adjacent to $m$, edge $bm$ is in the subtree. In the latter case node $m$ has major edge connected to $a$, that is we have switched major for minor edge. Because $m$ is not adjacent to $1$, edge $bm$ is not in the subtree. That is, when the plain edge is adjacent to the subtree we switch, otherwise we leave alone.

The remaining edges have unmodified evolution and the order tree is unchanges. These are all the moves required for SS operations on a subtree.
\end{proof}

\begin{proof}
{\bf Lemma \ref{TreeTopEdgeLemma}} We prove the result by induction.

For the tree with two nodes ($n=1$), we have two subtrees, both of which result in $b_s=1$ with an unchanged tree under SS operations, so $\sum_{\{s \in S(T):b_s=r\}}O(T_s)=2O(T)$ as required. We now assume the result is true for all such trees with $n$ nodes (or less) below the root. Now consider a tree with $n+1$ nodes below the root, such as Figure \ref{RemainderNodes}Ai. 

\begin{figure}[t!]
\centering
\includegraphics[height=90mm,width=140mm]{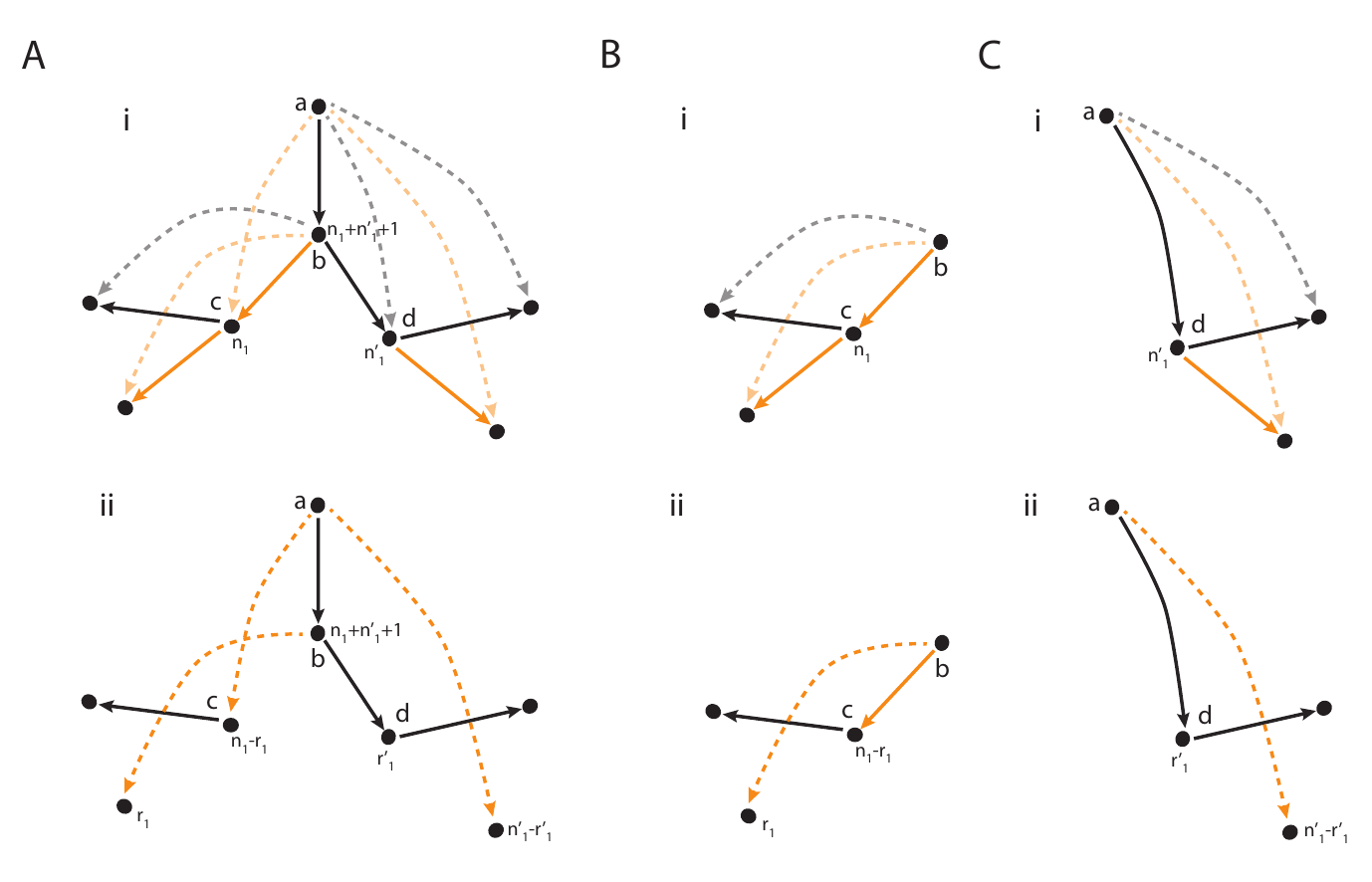}
\caption{In Ai we have a rooted tree with two branches. The bold edges indicate a subtree. In Bi and Ci we have the two branches as distinct trees along with their induced subtrees. In ii we have the corresponding trees after the ER operations. The node counts of the node $b$ below each root are indicated.}
\label{RemainderNodes}
\end{figure}

First consider the case $b_s = n+1$. There are two subtrees that leave the tree unchanged under SS operations. First is the empty subtree. The second subtree consists of the component of the tree composed of plain edges attached to the root. SS operations then leave the tree fixed and the order is the same. All other subtrees will send at least one edge to the root and reduce $b_s$. This gives us $\sum_{\{s \in S(T):b_s=n\}}=2O(T)$.

Now consider the case that $b_s = r+1 < n+1$. There can be two types of edges descending from the node below the root (node labelled $b$ in Figure \ref{RemainderNodes}). We can have a flipped edge, such as node $c$, or a plain edge such as node $d$. By Lemma \ref{MinorParent}, any children of flipped nodes $c$ must have minor edges connected to node $b$ below the root. No minor edges can touch the root. Conversely, any children of plain nodes such as $d$ must have minor edge attached to the root $a$, but not node $b$ below.

Suppose we have $i=1,2,...,I$ indexing flipped nodes $c_i$ adjacent to $b$ and $j=1,2,...,J$ indexing plain nodes $d_j$ adjacent to $b$. Any subtree $s$ of $T$ is going to result in a modified order tree such as Figure \ref{RemainderNodes}Aii following SS operations. We restrict attention to subtrees that result in a tree $T_s$ such that node $b$ has value $b_s = r+1$. Prior to SS operations flipped node $c_i$ has value $n_i$. After SS operations this becomes $n_i-r_i$, for some value $r_i$, with $r_i$ nodes now descending from node $b$. Prior to SS operations plain node $d_j$ has value $n'_j$. After SS operations this becomes some value $r'_j$ with $n'_j-r'_j$ nodes now descending from $b$. If $\sum_ir_i+\sum_jr'_j=r$ node $b$ then has value $r+1$.

Now the subtree $s$ can be split into subtrees $s_i$ and $s'_j$ that pass through nodes $c_i$ and $d_j$. We also subdivide tree $T$ according to nodes $c_i$ and $d_j$ into $T_i$ and $T'_j$ as follows. For flipped nodes we take nodes $c_i$, their descendants, node $b$, and all minors attached to these nodes (such as Figure \ref{RemainderNodes}Bi). For plain nodes we take nodes $d_j$, their descendants, node $a$, and all minors attached to these nodes (such as Figure \ref{RemainderNodes}Ci). The subtrees $s_i$ and $s_j$ can then induce SS operations on corresponding trees $T_i$ and $T'_j$.

It is convenient to define ratios $R(T_s)=\frac{O(T_s)}{O(T)}=\frac{\prod_k m_k}{\prod_k m_{k,s}}$, where $m_k$ and $m_{k,s}$ denote the node values of node $k$, as described in Lemma \ref{PosetReduction}, before and after the SS operations, respectively. Note that because the root node value (the total number of nodes in the tree) is unaffected by SS operations, $m_{root}$ and $m_{root,s}$ cancel and do not contribute to the sum $R(T_s)$.

We then find that,

\begin{align}
\begin{split}
\nonumber
\sum_{\{s \in S(T):b_s=r+1\}}R(T_s) =\sum_{\{r_1+...+r_R=r\}}\sum_{\{s_1 \in S(T_{1,s_1}):b_s=n_1-r_1\}}...\sum_{\{s_I \in S(T_{I,s_I}):b_s=n_I-r_I\}}. \\
 \hspace{10mm}\sum_{\{s'_1 \in S(T'_{1,s'_1}):b_s=r'_1\}}...\sum_{\{s'_J \in S(T'_{J,s'_J}):b_s=r'_J\}}\prod_{i=1}^I R(T_{i,s_i})\prod_{j=1}^J R(T'_{j,s'_j})\frac{1+\sum_i n_i+\sum_j n'_j}{1+r} \\
\end{split}
\end{align}

where we get a product of terms $R(T_s)$ from each subtree, along with the last term corresponding to node $b$. Because trees $T_i$ and $T'_j$ have less than $n+1$ nodes we can use the inductive hypothesis and this sum becomes,

\begin{align}
\begin{split}
\nonumber
&  \frac{1+\sum_i n_i + \sum_j n'_j}{1+r}\sum_{\{r_1+...+r_R=r\}}\sum_{\{s_1 \in S(T_1):b_s=n_1-r_1\}}R(T_{1,s_1})...\sum_{\{s_I \in S(T_I):b_s=n_I-r_I\}}R(T_{I,s_I}). \\
&     \hspace{10mm}\sum_{\{s'_1 \in S(T'_{1,s'_1}):b_s=r'_1\}}R(T_{1,s'_1})...\sum_{\{s'_J \in S(T'_{J,s'_J}):b_s=r'_1\}}R(T_{J,s'_J}) \\
& = \frac{1+n}{1+r}\sum_{\{r_1+...+r_R=r\}}\prod_{i=1}^I{^{n_i}C_{n_i-r_i}}\prod_{j=1}^J{^{n'_j}C_{r'_j}} = 
  \frac{1+n}{1+r}\sum_{\{r_1+...+r_R=r\}}\prod_{i=1}^I{^{n_i}C_{r_i}}\prod_{j=1}^J{^{n'_j}C_{r'_j}} \\
& = \frac{1+n}{1+r}{^nC_r} = {^{n+1}C_{r+1}}
\end{split}
\end{align}

Then substituting $R(T_s)=\frac{O(T_s)}{O(T)}$ gives the required result.
\end{proof}

\begin{proof}
{\bf Theorem \ref{LengthTheorem}}. We abuse notation throughout and equate random variables with their values. The required result can be demonstrated with induction. The initial distribution ($n=1$) of $L_1$ is the uniform distribution $U([0,2L])$, reflecting the uniform choice of the first breakpoint in $[0,L]$ prior to duplication, in agreement with the formula for $P(L_1)$. At each step of the BFB process we pick a breakpoint from $U([0,L_{n-1}])$ and double the length of the retained piece. We thus have $P(L_n|L_{n-1})=\frac{1}{2L_{n-1}}, 0\leq L_n \leq 2L_{n-1}$. Then assuming the form for $n-1$ we find that,
$P(L_n)= \int_{\frac{L_n}{2}}^{2^{n-1}L}P(L_n|L_{n-1})P(L_{n-1})dL_{n-1}$.
An integration by parts then gives the desired form.

Now $(L_n|L_{n-1}) \sim U([0,2L_{n-1}])$ gives us the Martingale property that $E_{L_n|L_{n-1}}(L_n)=L_{n-1}$, thus the initial value $E(L_1)=L$ tells us that the mean length of the BFB segment is $L$.

The variance $\int L_n^2P(L_n)dL-L^2$ follows from an integration by parts.
\end{proof}

\begin{proof}
{\bf Theorem \ref{SurvivalCriterion}}. This can be shown inductively. Initially, if the length of the structure produced by the $m^{\emph{th}}$ BFB cycle is $L_m$, then clearly the fold at the midpoint is a distance $\frac{L_m}{2}$ from either end after the $m^{\emph{th}}$ event. We then assume that all copies of the $m^{\emph{th}}$ BFB fold are at least a distance $\frac{L_m}{2}$ from either end of the structure prior to the $n^{\emph{th}}$ BFB (so $n>m$). One of two things can happen. Either the $n^{\emph{th}}$ fold is nearer to the ends than $\frac{L_m}{2}$ (so $L_n<L_m$), in which case all the copies of the $m^{\emph{th}}$ BFBs are deleted, or $L_n>L_m$ and some of them are duplicated, including a BFB nearest to the end, so the smallest distance $\frac{L_m}{2}$ is preserved, as required. 
\end{proof}

\begin{proof}
{\bf Theorem \ref{OrderTheorem}}. We first establish the formula for $W_k(x,y)$ by induction. We have initial value $W_1(x,y)=\int_x^{2y}\frac{dz}{z}=\log(2y)-\log(x)$ which matches the result. We assume true for $m$. Now we have $W_{m+1}(x,y)=\int_x^{2y}\frac{W_m(x,z)}{z}dz=\int_x^{2y}\sum_{j=0}^{m}a_{j+1}^{m}(x)\frac{\log^j(2^{m}z)}{z}dz$. Integration by parts gives $\int_x^{2y}\frac{\log^j(2^{m}z)}{z}dz=\frac{1}{j+1}(\log^{j+1}(2^{m+1}y)-\log^{j+1}(2^{m}x))$ so that $W_{k+1}(x,y)=\sum_{j=0}^{m}a_{j+1}^{m}(x)\frac{1}{j+1}(\log^{j+1}(2^{m+1}y)-\log^{j+1}(2^{m}x))$. Thus we find $a_1^{m+1}(x)=-\sum_{j=0}^{m}\frac{1}{j+1}a_{j+1}^{m}(x)\log^{j+1}(2^mx)$ and $a_{j+1}^{m+1}(x)=\frac{1}{j}a_{j}^{m}$, for $j \ge 1$, so that $a^{m+1}=B_{m+1}a^m$, as required.

Now if we have length $L_{n-1}$ prior to the $n^{th}$ BFB, then assuming the fold occurs uniformly along the length, the BFB duplication results in $P(L_n|L_{n-1})=\frac{1}{2L_{n-1}}$ with $0<L_n<2L_{n-1}$. The length sequence $L_n$ is also Markovian. Thus we can write $P(L_1,L_2,...,L_n)=\prod_{k=1}^nP(L_k|L_{k-1})=\frac{1}{2^{n}L\prod_{k=1}^{n-1}L_k}$ where $L_n<2L_{n-1}<...<2^{n-1}L_1<2^nL_0=2^nL$. 

The probability that the $k^\emph{th}$ of $N$ BFBs have minimum length $L_k=x$ is given by $M_{k,N}(x,L)=Pr(L_1 \ge x,...,L_{k-1} \ge x,L_k=x,L_{k+1} \ge x,...,L_N \ge x)=Pr(L_1,L_2,...,L_{k-1} \ge x,L_k=x)Pr(L_{k+1},...,L_N \ge x|L_k=x)$, where we have used the Markovian property of the length sequence.

The first term can be obtained by integrating the above density,

\begin{align}
\begin{split}
\nonumber
Pr(L_1,L_2,...,L_{k-1} \ge x,L_k=x) & =\frac{1}{2^{k}L}\int_x^{2L}\int_x^{2L_1}...\int_x^{2L_{k-2}}\frac{1}{L_1...L_{k-1}}dL_{k-1}...dL_1 \\
& =\frac{1}{2^{k}L}W_k(x,L)
\end{split}
\end{align}

The second term we similarly find as,

\begin{align}
\begin{split}
\nonumber
Pr(L_{k+1}, & L_{k+2},...,L_N \ge x |L_k=x)  =\frac{1}{2^{N-k}}\int_x^{2x}\int_x^{2L_k}...\int_x^{2L_{N-1}}\frac{1}{L_k....L_{N-1}}dL_N...dL_{k+1} \\
& =\frac{1}{2^{N-k}}\int_x^{2x}\int_x^{2L_k}...\int_x^{2L_{N-2}}\left[ \frac{2}{L_k....L_{N-2}}-\frac{1}{L_{k+1}....L_{N-1}}\right] dL_{N-1}...dL_{k+1} \\
& =\frac{1}{2^{N-k-1}}\int_x^{2x}\int_x^{2L_k}...\int_x^{2L_{N-2}}\frac{1}{L_k....L_{N-2}}dL_{N-1}...dL_{k+1}-\frac{1}{2^{N-k}}W_{N-k}(x,x) \\
& \vdots \\
& =1-\sum_{i=1}^{N-k}\frac{1}{2^i}W_i(x,x) \\
\end{split}
\end{align}

Putting these two terms together gives the required form.
\end{proof}

\begin{proof}
{\bf Corollary \ref{LengthCor}}v. The required integral can be recursively split as follows:
\begin{align}
\begin{split}
\nonumber
\frac{1}{2^dL_{i-1}} & \int_{L_i}^{2L_{i-1}}\int_{L_i}^{2l_1}...\int_{L_i}^{2l_{d-1}}\frac{1}{l_1....l_{d-1}}dl_d...dl_1 \\
& =\frac{1}{2^dL_{i-1}}\int_{L_i}^{2L_{i-1}}\int_{L_i}^{2l_1}...\int_{L_i}^{2l_{d-2}}\left[ \frac{2}{l_1....l_{d-2}}-\frac{L_i}{l_1....l_{d-1}}\right] dl_{d-1}...dl_1 \\
& =\frac{1}{2^{d-1}L_{i-1}}\int_{L_i}^{2L_{i-1}}\int_{L_i}^{2l_1}...\int_{L_i}^{2l_{d-2}}\frac{1}{l_1....l_{d-2}}dl_{d-1}...dl_1-\frac{L_i}{2^{d}L_{i-1}}W_{d-1}(L_i,L_{i-1}) \\
& \vdots \\
& =1-\frac{L_i}{2L_{i-1}}\sum_{k=0}^{d-1}\frac{1}{2^k}W_k(L_i,L_{i-1}) \\
\end{split}
\end{align}

\end{proof}

\footnotesize{

}

\end{document}